\documentclass[10pt,acmsmall]{acmart}
 
 \setcopyright{acmlicensed}
 \acmJournal{POMACS}
 \acmYear{2023} \acmVolume{7} \acmNumber{3} \acmArticle{55} \acmMonth{12} \acmPrice{15.00}\acmDOI{10.1145/3626786}
 
\usepackage{algorithmic}
\usepackage{graphicx}
\usepackage{textcomp}
\usepackage{xcolor}
\usepackage{colortbl}
\usepackage{multirow}
\usepackage{wrapfig}
\usepackage{mathtools}
\usepackage{adjustbox}
\usepackage{subfigure}

\newcommand{\ie}{{\em i.e., }}
\newcommand{\eg}{{\em e.g., }}

\begin{document}

\title[MetaVRadar: Measuring Metaverse Virtual Reality Network Activity]{MetaVRadar: Measuring Metaverse Virtual Reality\\ Network Activity}

\author{Minzhao Lyu}
\affiliation{
	\institution{University of New South Wales}
	\city{Sydney}
	\state{NSW}
	\country{Australia}
}
\email{minzhao.lyu@unsw.edu.au}

\author{Rahul Dev Tripathi}
\affiliation{
	\institution{University of New South Wales}
	\city{Sydney}
	\state{NSW}
	\country{Australia}
}
\email{r.tripathi@student.unsw.edu.au}

\author{Vijay Sivaraman}
\affiliation{
	\institution{University of New South Wales}
	\city{Sydney}
	\state{NSW}
	\country{Australia}
}
\email{vijay@unsw.edu.au}

\begin{abstract}
	The ``metaverse'', wherein users can enter virtual worlds to work, study, play, shop, socialize, and entertain, is fast becoming a reality, attracting billions of dollars in investment from companies such as Meta, Microsoft, and Clipo Labs. Further, virtual reality (VR) headsets from entities like Oculus, HTC, and Microsoft are rapidly maturing to provide fully immersive experiences to metaverse users. However, little is known about the network dynamics of metaverse VR applications in terms of service domains, flow counts, traffic rates and volumes, content location and latency, etc., which are needed to make telecommunications network infrastructure ``metaverse ready'' to support superlative user experience in the coming future.
	
	This paper is an empirical measurement study of metaverse VR network behavior aimed at helping telecommunications network operators better provision and manage the network to ensure good user experience. Using illustrative hour-long network traces of metaverse sessions on the Oculus VR headset, we first develop a categorization of user activity into distinct states ranging from login home to streetwalking and event attendance to asset trading, and undertake a detailed analysis of network traffic per state, identifying unique service domains, protocols, flow profiles, and volumetric patterns, thereby highlighting the vastly more complex nature of a metaverse session compared to streaming video or gaming. Armed with the network behavioral profiles, our second contribution develops a real-time method \textit{MetaVRadar} to detect metaverse session and classify the user activity state leveraging formalized flow signatures and volumetric attributes. Our third contribution practically implements \textit{MetaVRadar}, evaluates its accuracy in our lab environment, and demonstrates its usability in a large university network so operators can better monitor and plan resources to support requisite metaverse user experience.
	
\end{abstract}




\maketitle

\section{Introduction}
% Metaverse concept and growth
The term ``metaverse'' embodies the concept of an immersive virtual world for social interaction. Users, generally in the form of an avatar, immerse themselves into a 3D world and interact with animate and inanimate objects without being limited by aspects like distance, ability, and strength \cite{MBarredaAngelesCHB2022, FGenzARVRCG2021,DSaffoCHI2021}. Importantly, unlike other single-purpose networked applications such as online gaming and social media, the metaverse aims to provide users a wholistic experience encompassing their daily activities, such as working in the office, conducting meetings, socializing with friends, attending classes, shopping, playing games, participating in concerts and ceremonies, etc. \cite{ConsumerThought2021,ABhatia2022,YECCP2022}. To this end, metaverse platforms allows developers to create and operate their own events -- recent examples include a wedding ceremony \cite{SKurutz2021}, a charter school \cite{ANChulain2022}, and a university campus \cite{BerkeleyNews2020}.

% Accompanied by rise of VR
Early versions of the metaverse, embodied in platforms such as SecondLife, Minecraft, Sandbox, and Decentraland, operated on 2D Personal Computer screens, and lacked the ability to give users immersive experiences when interacting with the virtual environment. Recent advances in virtual reality (VR) wearable hardware, from vendors such as Oculus, HTC, Microsoft, and Apple have enabled true immersive experiences that provide 3D visualization along with body movement sensing so that users can perceive and interact with the virtual environment in a natural way. The Economist predicts that the market value of VR headsets will grow from \$5B in 2021 to \$12B in 2026 \cite{Economist2022}, and VR metaverse will create new experiences spanning all sectors including entertainment, education, healthcare, and remote work. Indeed, companies like Meta, Google, Microsoft, and ByteDance, are investing heavily into building their VR ecosystems.

% Telco interest in Metaverse VR
The telecommunications industry is becoming increasingly keen to tap into the opportunity of VR metaverse, and starting to explore how network functionality and performance need to evolve to support immersive services. The Telecom Infra Project (TIP) has launched a new project group in 2022 to address ``metaverse readiness'' \cite{AWeissberger2022} whose primary objective is to {\em accelerate the development of solutions and architectures that enhance network readiness to support metaverse experiences}, addressing aspects such as greater agility, programmability, performance, and reliability in both fixed and mobile networks. Telecommunications operators such as T-Mobile and Telefonica have thrown their weight behind this, while others like SingTel and SK Telcom are also talking of developing offerings with high-speed connectivity and ultra-low latency that enable enhanced metaverse experience \cite{Singtel}.

% Telcos do not know what to optimize
To enhance the metaverse VR experience, telecommunications operators however need to go beyond a simplistic view focused just on access bandwidth and latency. They need to possess a thorough understanding of how the application behaves on the network, including: (a) which servers and locations does the application access data from? (b) what are the duration, volumes, and rates of the various data transfers; and (c) which segments of the session are most relevant to user experience? As we will soon show, unlike a streaming video or gaming session that involves fairly predictable content exchange with just one or a few servers \cite{HHGharakheiliTNSM2019,SCMadanapalliPAM2022}, a metaverse session can be tremendously more complex, involving tens of autonomous systems, hundreds of domains, and thousands of traffic flows. Without fully understanding the application network dynamics, telecommunications providers will be unable to monitor the performance of the various component blocks, such as user-generated computational elements \cite{LLeeArXiv2021Dec}, edge compute \cite{HNingArXiv2021}, and distributed token systems \cite{QYangArXiv2022}. This will limit their ability to focus their efforts on optimizing the components that impact user experience the most.

% The gap we address
This paper addresses this gap by conducting a comprehensive measurement study of metaverse VR network behavior. The prior measurement studies we are aware of \cite{cheng2022are,MetaverseQoE} focus on metaverse bandwidth, latency, and computational needs, with the view (for metaverse developers) to optimizing the platforms to scale to a large number of users. Our study by contrast is aimed at \textbf{telecommunications network operators}, helping them understand metaverse usage and experience, so they can make better planning and provisioning decisions.
%our contributions
Specifically, we study the network anatomy of four popular Metaverses (Multiverse, VRChat, Rec Room, and AltspaceVR) on a dominant VR platform (Oculus), and make the following three contributions. 

Our \textbf{first contribution} (\S\ref{sec:Development}) highlights the complex behavior of metaverse VR using an illustrative trace, and makes the analysis tractable by discretizing user activity into a small number of states that are categorized into six distinct types. We discuss how network traffic patterns differ across user activity states (such as in domains accessed, flows established, and volumes transferred), identify network factors that most impact user experience under each type of state,  and discuss performance requirements for telecommunications network operators to accommodate metaverse users.

Our \textbf{second contribution} (\S\ref{sec:characteristics}) develops a method \textit{MetaVRadar} to detect metaverse VR session and classify user activity state via real-time analysis of network traffic. Using our analytical insights on domains accessed and packet payload sizes, we develop robust signatures to detect active metaverse sessions; and using tens of packet- and flow-level volumetric attributes extracted periodically (\eg every few seconds) we develop stateful machine-learning models to make a determination on user activity states with high confidence. 

Our \textbf{third contribution} (\S\ref{sec:detectMetaApp}) implements a prototype of \text{MetaVRadar}, and evaluates its accuracy in metaverse session detection and user activity state classification in our lab environment. We then deploy it in a large-sized University network, and make observations on metaverse VR activity in the wild. We find instances where metaverse sessions are spread across half a dozen ASes, with user-created events dominating the volume consumption and experiencing poor latencies from one AS. Our study will help telecommunications operators use such insights to better tune their network resources (bandwidth, latency, caching, routing, etc.) to improve user experience and make their infrastructure ``metaverse ready''.

\begin{table*}[!t]
	\caption{Five popular metaverse virtual reality (VR) applications and their current specifications.}
	\centering
	\small
	\begin{tabular}{|p{1.465cm}|p{1.8cm}|p{2.3cm}|p{1.7cm}|p{1.22cm}|p{3.2cm}|}
		\hline
		\rowcolor[rgb]{ .906,  .902,  .902}  \textbf{Metaverse}            &\textbf{Developer}               & \textbf{Supported Platform}                                                          & \textbf{Currency System} & \textbf{Event Venue}  & \textbf{User Popularity} \\ \hline
		Horizon Worlds  \cite{HorizonWorlds} & Meta Platform               & \textbf{\color{blue}OculusVR}                                                        & Centralized token system       & Provider approval    & $\sim$300K monthly active users in Feb 2022 \cite{DHeaney}           \\ \hline
		Multiverse \cite{Multiverse}    & Clipo Labs              & \textbf{\color{blue}OculusVR}                                     & Supporting NFT    &  Virtual property &          Not reported        \\ \hline
		VRChat \cite{VRChat} & VRChat Inc.             & \textbf{\color{blue}OculusVR}, HTC Vive VR                                       & Supporting NFT         & Provider approval &     $\sim$24K max concurrent online users in Jul 2021 \cite{ARoy}           \\ \hline
		RecRoom \cite{RecRoom}      & RecRoom Inc.           & \textbf{\color{blue}OculusVR}, PlayStation VR                        & Supporting NFT      & Provider approval  &   $>$1M monthly active users in Jan 2021 \cite{WJAu}                  \\ \hline
		AltspaceVR \cite{AltspaceVR}    & Microsoft               & \textbf{\color{blue}OculusVR}, HTC VR, Windows MR & Not available            & Provider approval        &         Not reported            \\ \hline
	\end{tabular}
	\label{tab:7MetaverseApps}
\end{table*}

\section{Background \& Preliminaries}\label{sec:Dataset}
We briefly review the popular metaverse applications and virtual reality (VR) platforms in the market today (\S\ref{sec:PopularApp}). We then describe our Oculus VR lab setup for immersion into four popular metaverse VR applications and traffic trace collection (\S\ref{sec:labSetup}). We finally describe a representative walk-through of a typical metaverse VR session (\S\ref{sec:WalkExample}). 

\subsection{VR Platforms and Metaverse Applications}\label{sec:PopularApp}
Among VR headsets in the market today, the Oculus VR platform, owned by Meta (\ie Facebook) is the most dominant, accounting for nearly 80\% of the market. The Oculus Rift model needs to be tethered to a graphics PC, whereas the more modern Quest is stand-alone. Other significant VR headset manufacturers include HTC Vive (models include Focus, Pro, and Cosmos), Microsoft HoloLens, Lenovo Explorer, and Dell Visor, with Playstation VR, Pico VR, and Apple Vision Pro joining this fast-expanding market recently.

Table~\ref{tab:7MetaverseApps} lists four popular metaverse applications emerging in the market. {\em Horizon Worlds} is Meta's flagship metaverse application, available exclusively on the Oculus VR platform. It reported about 300K monthly active users in Feb 2022 \cite{DHeaney}, and uses an in-app centralized token currency. Users are allowed to create their own events (\eg for socializing, working, education, and entertainment), and require approval from the operator before being made accessible to other users. Horizon Worlds is currently limited to the North America market.
{\em Multiverse} is another popular metaverse, developed and operated by unicorn startup Clipo Labs. It allows users to buy, rent, and trade virtual properties (\eg residential apartments, retail shops, concert halls), where they can host events such as parties, sales, and ceremonies. Multiverse claims to be one of the most versatile VR application, allowing third parties to support transactions via NFTs \& cryptocurrencies.
Other popular VR metaverses include VRChat, Rec Room, and AltspaceVR (shut down in March 2023) with attributes as listed in Table~\ref{tab:7MetaverseApps}. All the listed metaverse applications are available on the Oculus VR platform.

\subsection{Measurement Setup and Data Collection}\label{sec:labSetup}
\begin{wrapfigure}{L}{0.55\textwidth}
	\includegraphics[width=0.55\textwidth]{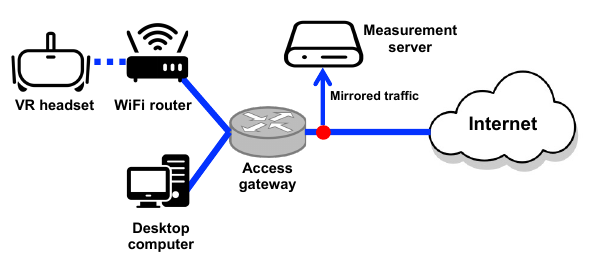}
	\caption{Our lab measurement setup.}
	\label{fig:MeasurementSetup}
\end{wrapfigure}

The laboratory measurement infrastructure to collect traffic traces of metaverse VR sessions is shown in Fig.~\ref{fig:MeasurementSetup}. A VR headset (Oculus Quest 2) is connected to a WiFi router in the lab which connects to the campus network to reach the Internet. A desktop computer is also connected to the lab network to generate non-metaverse traffic, needed to evaluate our metaverse detection algorithms. The network traffic between the lab and the Internet is mirrored to a measurement server. Filters are provisioned for selective packet captures (\ie PCAPs) pertaining to the end-points of interest (\eg the WiFi router to which the VR headset connects). 
Two of the authors participated in the data collection process.
During each metaverse session, the timeline and description of activities are logged as the example in Table~\ref{tab:examplePlay}, indicating what the user was doing over each period within the metaverse. The collected lab traffic traces (\ie PCAPs) and user activity logs are used for ground-truth traffic analytics (\S\ref{sec:Dataset} and \S\ref{sec:Development}), development of \textit{MetaVRadar} (\S\ref{sec:characteristics}), and in-lab evaluation (\S\ref{sec:labEval}).

In addition to the lab setup, our university campus traffic is tapped\footnote{We have obtained ethics clearance (UNSW Human Research Ethics Advisory Panel approval number HC211007) to analyze campus traffic for AR/VR applications without identifying user identities. In our campus deployment, we report aggregated metaverse statistics behind WiFi gateways serving dormitory residents.} at the University's egress gateway to the Internet and mirrored to our real-time \textit{MetaVRadar} prototype operated on a Ubuntu server. The server is equipped with dual processor 16-core Intel Xeon 2.10GHz CPU, 64GB RAM, and two 10Gbps NICs for upstream and downstream mirrored traffic, respectively. Flow statistics of metaverse sessions are logged for our campus network deployment insights (\S\ref{sec:uniEval}).

We note that our method uses volumetric profiles of flows and metadata extracted from layer-3 (\ie IP) headers, layer-4 (\ie UDP or TCP) headers, and Server Name Indication (SNI) from TLS client hello messages. While the metadata remains unchanged during packet routing between clients and servers, slight variations in volumetric profiles can occur among measurement vantage points due to arrival times. Thus, as will be discussed in \S\ref{sec:discussionOnLimitation}, specific models should be trained for their deployment environment to ensure optimal performance.

\begin{figure*}[t!]
	\includegraphics[width=1\textwidth]{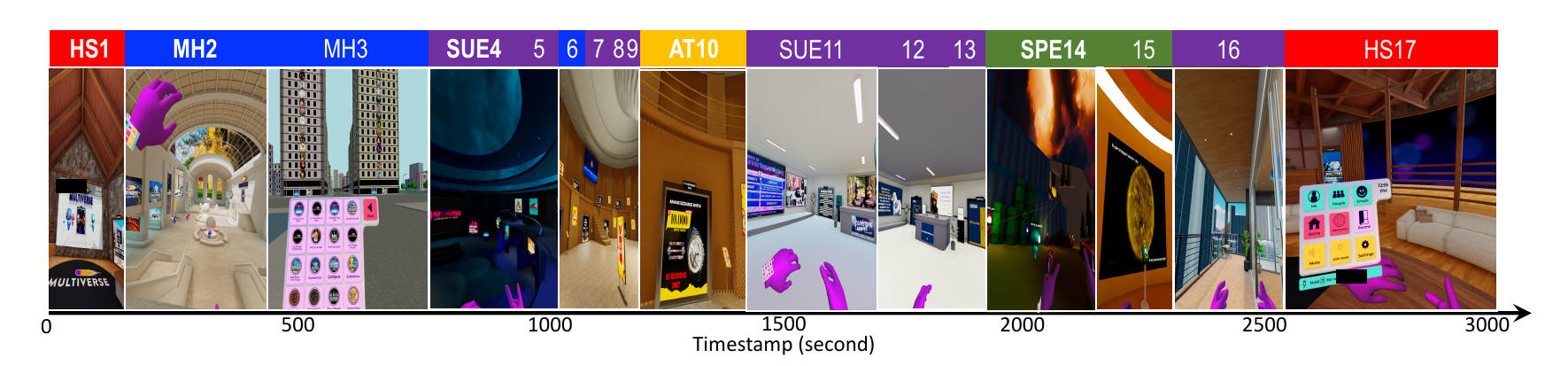}
	\vspace{-5mm}
	\caption{Screenshots of example user activities and their states in one Metaverse (Multiverse) session.}
	\label{fig:caseStudy}
\end{figure*}

\subsection{A Representative Metaverse VR Walk-Through}\label{sec:WalkExample}
To relate metaverse user activity with network activity, we begin with an illustrative 45-minute session in one of the most popular metaverses -- the Multiverse application on the Oculus Quest 2 VR headset. Screenshots illustrate the activities in Fig.~\ref{fig:caseStudy}, with associated timelines and description provided in Table~\ref{tab:examplePlay}. Once we put on the headset and open the Multiverse application console, we start the login and initialization processes in our home space {\bf (HS)} (HS1, 0--159s), which allows us to remain online without being involved in social activities. We then exit our home and enter the Main Hub {\bf (MH)}, which could be a public interchange station (MH2, 160--457s) or the public street (MH3, 458--788s), from where we can access portals to other events hosted either by the Multiverse provider or in other users' virtual properties. It can be seen in MH3 that each floor of the building is owned by a different user, each of whom can host freely accessible, invitation-only, or paid commercial events within their properties. We can explore the events by entering the building or checking the interactive menu for nearby properties. Unlike Multiverse, the VRChat metaverse has not implemented social settings of MH; instead, it uses pop-up menus for this purpose.

From the public street, we visit several separate user-created events {\bf (SUE)}, such as a movie theater and its associated media room (SUE4 -- 5, 789 -- 1059s) in a commercial property, and a user-created shopping mall (SUE7--9, 1113--1224s) displaying assets on various levels (transitions are made though the main hub MH6). We browse through a store selling watches, and can choose to perform asset trading {\bf (AT)} using digital currencies -- in AT10 (1225--1449s) we can see pop-ups showing details of the items available for trading with links to complete the purchase.

\begin{table}[!t]
	\caption{Description of user activities in Fig.~\ref{fig:caseStudy}.}
	\small
	\begin{tabular}{|l|l|p{9cm}|}
		\hline
		\rowcolor[rgb]{ .906,  .902,  .902} 	\textbf{Label}  & \textbf{Timeline} & \textbf{Description} \\ \hline
		\textbf{\color{red}HS1}  & 0 -- 159s      &   Login to the metaverse and enter my private home space         \\ \hline
		\textbf{\color{blue}MH2}    & 160 -- 447s       &  Enter the grand central (exchange station)    \\ \hline
		\textbf{\color{blue}MH3}    & 448 -- 788s      &       Enter the public street hosting portals to other events     \\ \hline
		\textbf{\color{purple}SUE4}   & 789 -- 973s      &      Enter a user-created theater from the public street       \\ \hline
		\textbf{\color{purple}SUE5}   & 974 -- 1059s        &         Enter a media room in the theater \\ \hline
		\textbf{\color{blue}MH6}    & 1060 -- 1112s      &          Go back to the public street   \\ \hline
		\textbf{\color{purple}SUE7}   & 1113 -- 1164s      &         Enter a user-created shopping mall    \\ \hline
		\textbf{\color{purple}SUE8}   & 1165 -- 1195s      &          Go to a different level of the mall  \\ \hline
		\textbf{\color{purple}SUE9}   & 1196 -- 1224s   &           Go to a different level of the mall   \\ \hline
		\textbf{\color{brown}AT10}   & 1225 -- 1449s    &           Browse and purchase watches selling in the mall \\ \hline
		\textbf{\color{purple}SUE11}  & 1450 -- 1719s &         Teleport to a business showroom via the menu of nearby events   \\ \hline
		\textbf{\color{purple}SUE12}  & 1720 -- 1869s &          Teleport to another business show room   \\ \hline
		\textbf{\color{purple}SUE13}  & 1870 -- 1944s &            Teleport to another business show room \\ \hline
		\textbf{\color{olive}SPE14}  & 1945 -- 2219s  &           Play a provider-created multi-player shooting game   \\ \hline
		\textbf{\color{olive}SPE15}  & 2220 -- 2332s &           Enter a provider-created NASA space gallery  \\ \hline
		\textbf{\color{purple}SUE16}  & 2333 -- 2559s    &     Visit a friend's property in an residential apartment   \\ \hline
		\textbf{\color{red}HS17} & 2560 -- 2700s      &      Teleport back to my private home space       \\ \hline
	\end{tabular}
	\label{tab:examplePlay}
\end{table}

We then visit a business showroom from a not-for-profit organization helping people with addiction, and a company providing VR-based education (SUE11--13, 1450--1944s). After that, we try separate provider-created events {\bf (SPE)} that are operated by the Multiverse provider, specifically a shooting game (SPE14, 1945--2219s) and a space gallery (SPE15, 2220--2332s). We then visit a friend's place in a residential apartment to attend an invitation-only balcony party (SUE16, 2333-2559s), before teleporting back to our private space (HS17, 2560--2700s).

The various activities including entering, walking, socializing, playing, shopping, educating, and entertaining captured in the 45-minute trace above are typical of metaverse behavior, and we will continue to use this as a running example throughout the paper. Later in \S\ref{sec:Development}, we categorize the activities into distinct states (HS, MH, SUE, SPE, AT) in order to model user behavior and ease the development of network activity profiles. A sixth state, content creation {\bf (CC)}, that allows users to create their own digital assets and build virtual environments, will also be considered, but has been excluded from the illustration above as it is done outside the metaverse console for Multiverse.

\section{Metaverse VR Network Traffic Characteristics}\label{sec:Development}
In this section we describe the anatomy of metaverse VR network traffic, illustrating its complexity and dependence on user activity.
We spent tens of hours extensively immersed in each of the metaverse VR applications listed in Table~\ref{tab:7MetaverseApps} (except for Horizon Worlds, which is not available in our geography), maintaining logs of our activities (\ie timestamp and description) and correlating it with network packet captures. We found that the user activity states introduced above for the illustrative metaverse trace are broadly applicable across the various immersive worlds (with some differences, which we will highlight where appropriate).
With the goal of achieving comprehensive coverage of network traffic characteristics influenced by user movements, we engaged in a diverse range of available body movements and gesture interactions within each event, such as waving in transit stations, walking on public streets, joining in conversations with others, dancing at parties, and participating in combat scenarios where we targeted enemies and evaded attacks.

%LI
\textbf{Distinct user activity states:} All metaverse VR applications have a home space (\textbf{HS}), where the user enters upon login, performs initialization, and can have private time in the virtual world. In most metaverses (except VRchat), the user then enters the main hub (\textbf{MH}), which could be a street, plaza, or station, where they can walk around and interact with users and events. VRchat is an exception in that it does not have a main hub, and instead uses pop-up menus. All the metaverses considered allow third party separate user-created event (\textbf{SUE}), wherein users can attends social events (\eg family party, business meeting, online educational class, and game) created and hosted by others. In addition Multiverse operates separate provider-created events (\textbf{SPE}), such as games, chatrooms, and virtual exhibitions built and directly operated by the metaverse provider. Multiverse and Rec Room support asset trading (\textbf{AT}), where users can browse and exchange any assets such as NFTs and currencies. Content creation (\textbf{CC}) is generally done by users outside of the metaverse application itself, except Rec Room that allows in-app content creation.

\begin{figure*}[t!]
%	\vspace{-7mm}
	\mbox{
	%	\hspace{-2mm}
		\subfigure[Top ranked service domains.]{
			{\includegraphics[width=0.4\textwidth]{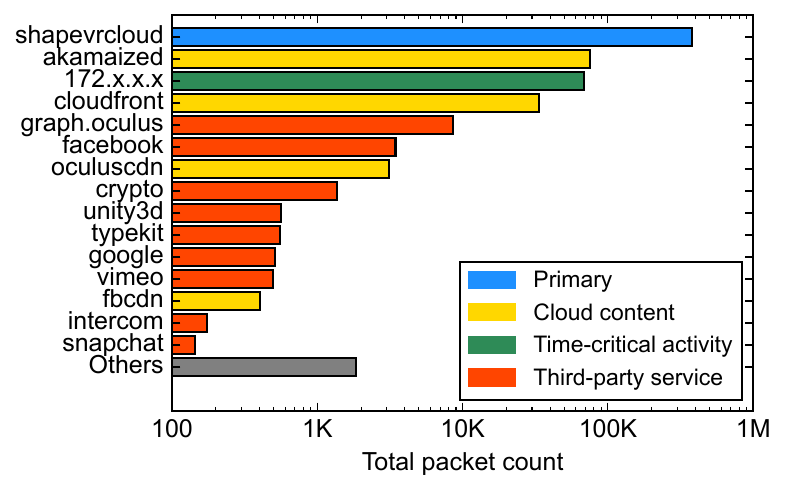}}\quad
			\label{fig:Multiverse2Domain}
		}
%		\hspace{-3mm}
		\subfigure[Flow distribution of our example session in Table~\ref{tab:examplePlay}.]{
			{\includegraphics[width=0.45\textwidth]{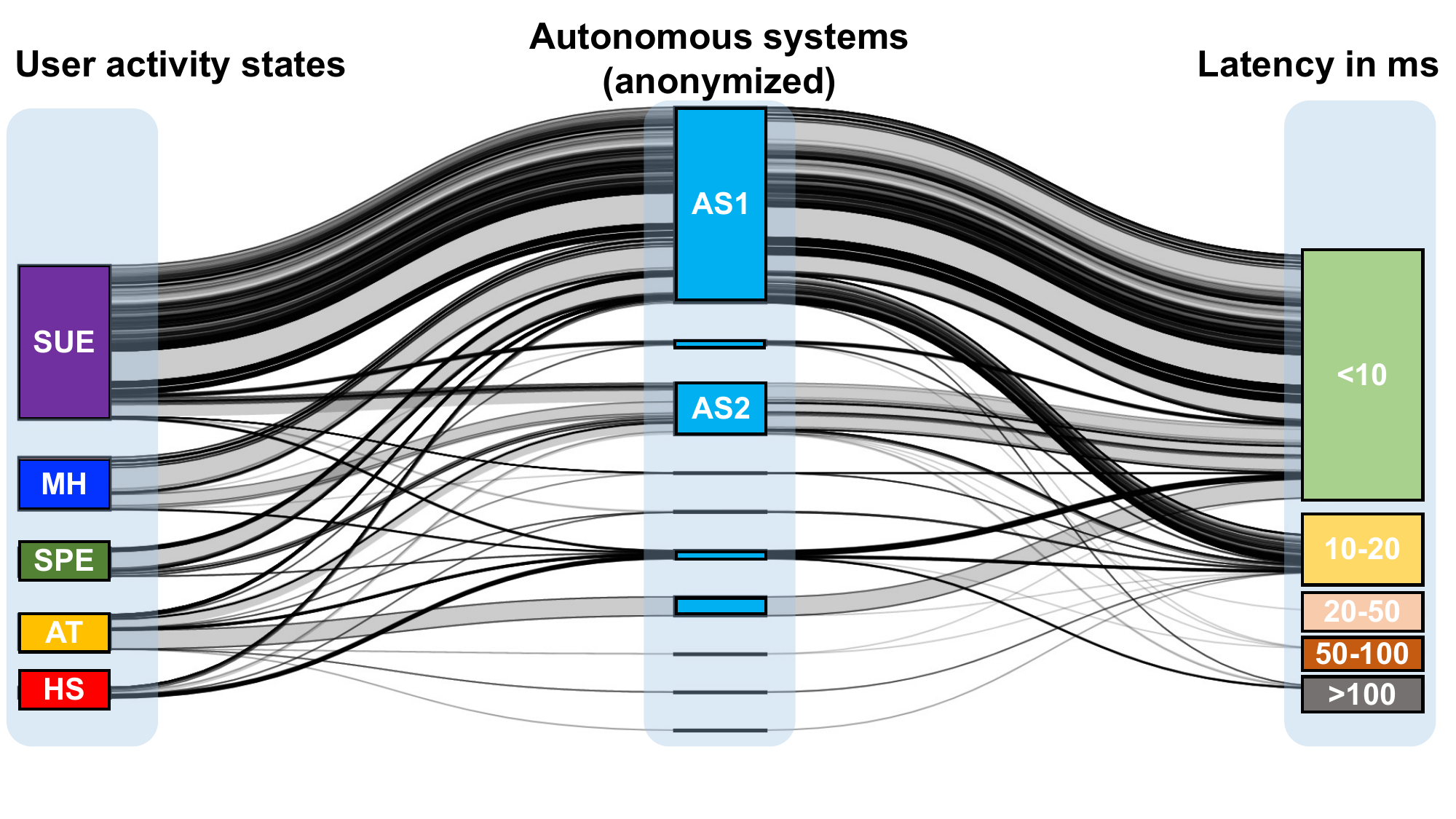}}\quad
			\label{fig:Sankey}
		}
	}
%	\vspace{-5mm}
	\caption{Complexity of metaverse user network activities in our representative walk-through: (a) service domains being accessed ranked by total packet count; and (b) distribution of flows across different user activity states, autonomous systems, and latency ranges.}
	\label{fig:categorizeUserActivity}
%	\vspace{-5mm}
\end{figure*}

\textbf{Service domains accessed:} To study the complex network activity within a metaverse session, we find it convenient to decompose it by the user activity states as categorized above. Further, since each metaverse session can interact with a large number of domains, to make the study tractable we categorize the domains into four classes based on ownership and purpose: \textbf{primary} domains are those directly operated by the metaverse application developer; \textbf{time-critical activity} domains are those providing real-time synchronization of user movement and activity; \textbf{cloud content} domains are general-purpose CDNs that for example serve video streams; and \textbf{third-party service} domains are for non-metaverse services such as advertisements and social media content. Based on the provided definition, we can assign tags to each accessed domain in a metaverse session using automated scripts. For example, service domains with domain names like \textit{vrchat.net} were labeled as the primary domain for VRChat, while \textit{oculuscdn.com} and \textit{fbcdn.com}, which are generic cloud servers, were labeled as cloud content domains. We also manually checked the registry information of the labeled domains to ensure that they were all correctly tagged.

With the user activity states and service domain categorizations in place, we first present some high-level characteristics of the traffic collected for the representative walk-through of Multiverse presented earlier. Overall, the session accessed 87 service domains. In Fig.~\ref{fig:Multiverse2Domain} we show the top 15 domains ranked by their packet counts, color-coded to represent their categories. Not surprisingly, the majority of packets are exchanged with the primary Multiverse domain {\em shapevrcloud}, which serves the environment (buildings, rooms, streets, etc.) information as the user moves. Cloud content domains {\em cloudfront} and {\em oculuscdn} seem to serve VR platform related images and advertisements, {\em akamaized} serves the video watched by the user in the theater in SUE5, and {\em fbcdn} provides the social media feeds. Servers in the {\em 172.x.x.x} subnet manage time-critical (UDP) activity to update user location and gesture as they move around in the metaverse. Third party services such as {\em unity3d, crypto}, etc. provide graphics, access to third-party businesses, and crypto services. As can be seen, the number of domains involved in serving a metaverse VR session is significantly higher compared to streaming video (which typically involve a single CDN domain serving video content) or gaming (which generally consist of less than ten prefixes within a single domain in a typical match-play game \cite{SCMadanapalliPAM2022}) sessions. The contrast highlights the challenges that operators will encounter when it comes to optimizing and troubleshooting their networks for metaverse.

Utilizing the Sankey diagram presented in Fig.~\ref{fig:Sankey}, we visualize the network flow distribution within the metaverse based on user activity states (nodes in the left column). Additionally, we highlight the autonomous systems (AS) from which the traffic originates (nodes in the middle column), and provide information on the measured latency categorized into five groups (nodes in the right column). Each line in the Sankey diagram represents a flow, with the width of the line proportional to the packet count. This description also applies to the individual Sankey diagrams for SUE, SPE, and AT, which are subsequently illustrated in Fig.~\ref{fig:flowDistributionSUESPE}.
We note that a majority of traffic corresponds to separate user-created events (SUE), since that is where users are expected to engage the most socially. Further, the latency for some of the content in SUE and asset trading (AT) is very high (even exceeding 100ms), as they are from third-parties and may not be from the local cache/CDN. Network operators might therefore want to selectively optimize their routing paths or caching strategies depending on the user activity state.

In what follows, we discuss the network traffic characteristics of each identified user activity state in details.

\subsection{Home Space (HS)}\label{sec:HS}
HS is where a user enters Multiverse. We analyze the network dynamics in terms of flow setup sequence, the domains that they communicate with, and the rates/volumes within these flows. The walk-through shown in Fig.~\ref{fig:caseStudy} will be used as the primary illustrative example, and we will highlight different behaviors in other metaverses.

\begin{figure*}[t!]
	\begin{center}
		\mbox{
			\hspace{-4mm}
			\subfigure[Flow span.]{
				{\includegraphics[width=\textwidth]{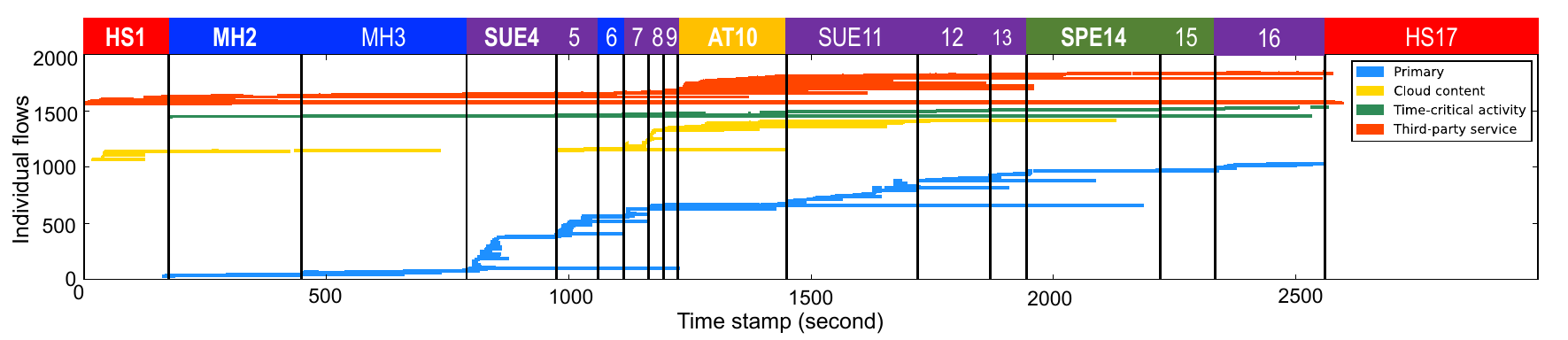}}\quad
				\label{fig:flowProfile}
			}		
		}
		\mbox{
			\hspace{-4mm}
			\subfigure[TCP packet rate.]{
				{\includegraphics[width=\textwidth]{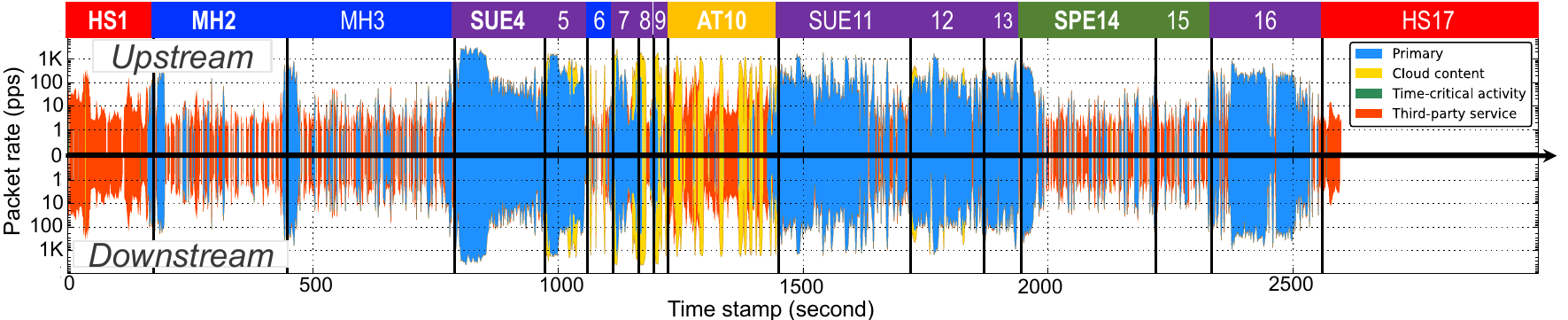}}\quad
				\label{fig:TCPPkt}
			}
		}
		\mbox{
			\hspace{-4mm}
			\subfigure[UDP packet rate.]{
				{\includegraphics[width=\textwidth]{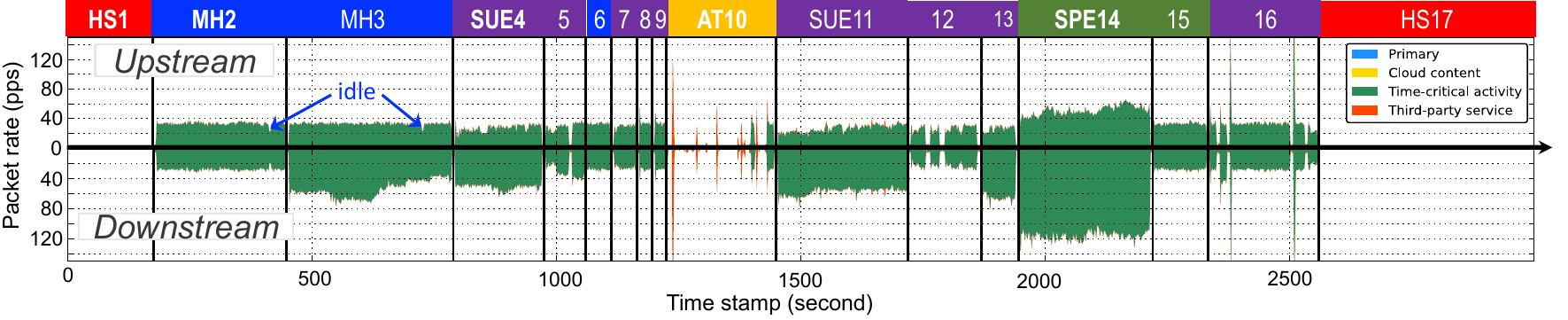}}\quad
				\label{fig:UDPPkt}
			}
		}
		\caption{Flow span and volumetric patterns of our one example Multiverse session consisting of 17 user activities.}
		\label{fig:volPattern}
	\end{center}
\end{figure*}

Fig.~\ref{fig:flowProfile} shows the span of the individual flows that are active during the metaverse session. Each line represents a 5-tuple, and is color-coded based on the type of domain (primary, time-critical activity, cloud content, or third-party service). We can see that in HS1, most of the flows are to third-party service (\eg \textit{graph.oculus}) and cloud content (\ie \textit{oculuscdn}) servers, which are used by the VR platform to fetch graphical data and exchange administrative information prior to commencing metaverse session. Once the metaverse activity commences, flows (seen in blue towards the end of HS1) are initiated to the primary metaverse domain (\ie \textit{shapevrcloud}) for Multiverse initialization including user login, account settings, and asset inventory loading. More specifically, six TCP flows are initiated to \textit{prod.shapevrcloud} followed by one TCP flow to \textit{prodblob.shapevrcloud}.

Fig.~\ref{fig:TCPPkt} and Fig.~\ref{fig:UDPPkt} show respectively the upstream/downstream TCP and UDP packet rates (measured every second) observed during the Multiverse session, color-coded as before based on the domain type (note that the y-axis of Fig.~\ref{fig:TCPPkt} uses a log scale for easier visualization). In the home state (HS) we observe that around 10-100 packets are exchanged per second with the Oculus servers, while the packet rate to the primary metaverse servers is over a hundred. There is no time-critical activity in this state.

We collected and analyzed similar traces on the other metaverse applications. For instance, AltSpaceVR (with its primary domain \textit{altvr}) in its home space (HS) creates flows towards three prefixes in the order \textit{config}, \textit{cdn-content-ingress}, and \textit{account}; VRChat has its sequential primary domain prefixes \textit{api}, \textit{pipeline}, \textit{assets}; and Rec Room (with its primary domain \textit{rec}) communicates with prefixes including \textit{api}, \textit{auth}, \textit{img}, \textit{cdn}, \textit{clubs}, and \textit{commerce}.

We note that all TCP flows are TLS-encrypted, and we are able to extract the SNI from the connection setup packets to identify the server the communication is with. We make note of the sequence of TCP connections and the payload lengths of the first few packets of each connection -- these will be used to develop our metaverse activity detection model in \S\ref{sec:trafficSignature}.

In terms of network requirements, HS is the least demanding state. The activity in HS is predominantly private without interaction with other users, so is not very sensitive to latency and jitter. Further, the service domains being accessed in HS are a small set, limited to the VR platform and the metaverse operator. The low to moderate bandwidth requirements in this state, coupled with the small number of domains the content is coming from, mean that this state should not be a major concern for network operators.

\subsection{Main Hub (MH)}\label{sec:MH}
After leaving the private home space, we entered a public interchange station and then a public street that serves as an immersive menu to other available events either hosted by individual users or directly created by the metaverse operator. It is also a social hub for all users who have not decided which event to go next. 

As shown in MH2, MH3 and MH6 of Fig.~\ref{fig:flowProfile}, there are active flows toward primary domains to fetch metaverse content such as surrounding environments. Given that MH events are developed by metaverse operators and most of the content (\eg city maps, building structures, and street layouts) has already been installed with the application console in the VR headset, only a small amount of dynamic content (\eg weather in the event and announcements from the developer) has to be downloaded from the primary domain at run-time. This leads to a fairly low packet rate and bandwidth usage for primary domain flows as shown in Fig.~\ref{fig:TCPPkt}, the exception being the beginning of each MH state whereby a bulk of metaverse content has to be downloaded.

In addition to primary flows that fetch event content via TCP at run-time, there are few flows toward time-critical activity domains (\ie green lines starting from MH2 in Fig.~\ref{fig:flowProfile}) synchronizing real-time user positions and avatar gestures in MH, as this state involves socialization among users. Those time-critical activity flows are purely via certain UDP ports \textit{5055}, \textit{5056}, or \textit{5058} which are commonly used by online gaming for the similar purpose. 
When we perform intensive motion updates like walking, dancing, and actively talking, as shown in Fig.~\ref{fig:UDPPkt}, the data exchange rates with time-critical activity domains during MH remain roughly constant at around 30pps in the upstream direction, whereas it drops to a lower rate of approximately 15pps when we stay idle as indicated by the blue arrows in Fig.~\ref{fig:UDPPkt}.
Downstream packet rates vary depending on the level of crowding (\ie number of active users and their motion updates) in the current event. For example, in MH3, we can observe variations in the downstream packet rates in Fig.~\ref{fig:UDPPkt} due to changes in the number of proximitous users. Similar observations can also be made for time-critical activity domains during SUE, SPE, and AT.

When exploring the interchange station (MH2) and public street (MH3), we also saw a number of posters, video advertisements, and third-party customized user avatars. Such content provided by commercial or residential users (\eg owners of a building) is often fetched from third-party service domains such as \textit{ads-twitter}, \textit{doubleclick}, \textit{redditstatic}, while some of the heavy ones are downloaded from dedicated cloud content providers like \textit{cloudfront} and \textit{fbcdn}. Therefore, we observe a non-negligible numbers of TCP packets toward those domains in the MH states.

For a network operator, guaranteeing low latency and stable jitter in both TCP and UDP is critical for user-perceived metaverse experience as application and third-party content are fetched from their respective service domains at run-time, which could be quite distributed in routing path. MH is not bandwidth demanding in TCP as most of the metaverse content has already been installed in the local console, while bandwidth availability for UDP has to be sufficient and stable due to constant synchronization of user activities in MH.

\subsection{Separate User-Created Event (SUE)}\label{sec:SUE}
In our representative walk-through, we entered several separate metaverse events such as theater (SUE4-5), shopping mall (SUE7-9), business showrooms (SUE11-13), and a friend's apartment (SUE16) created and hosted by individual commercial and residential users. 

\begin{figure*}[t!]
	\mbox{
		\hspace{-2mm}
		\subfigure[SUE.]{
			{\includegraphics[width=0.3\textwidth]{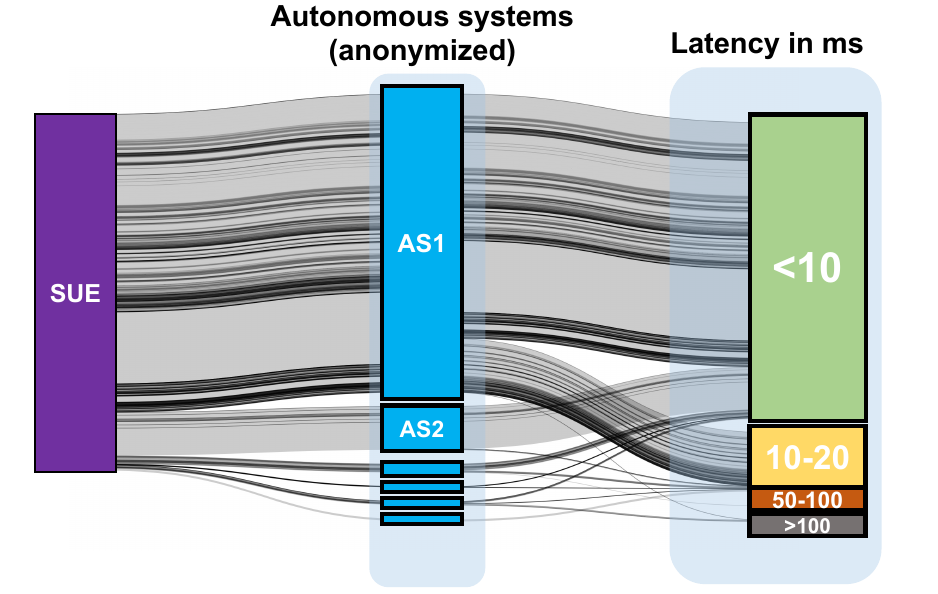}}\quad
			\label{fig:flowDistributionSUE}
		}
		\subfigure[SPE.]{
			{\includegraphics[width=0.3\textwidth]{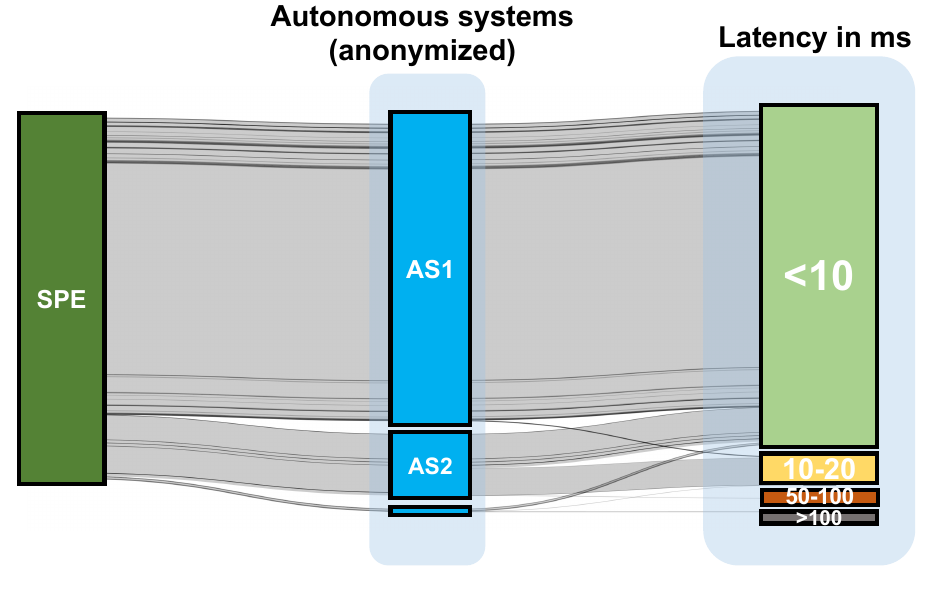}}\quad
			\label{fig:flowDistributionSPE}
		}
\subfigure[AT.]{
	{\includegraphics[width=0.3\textwidth]{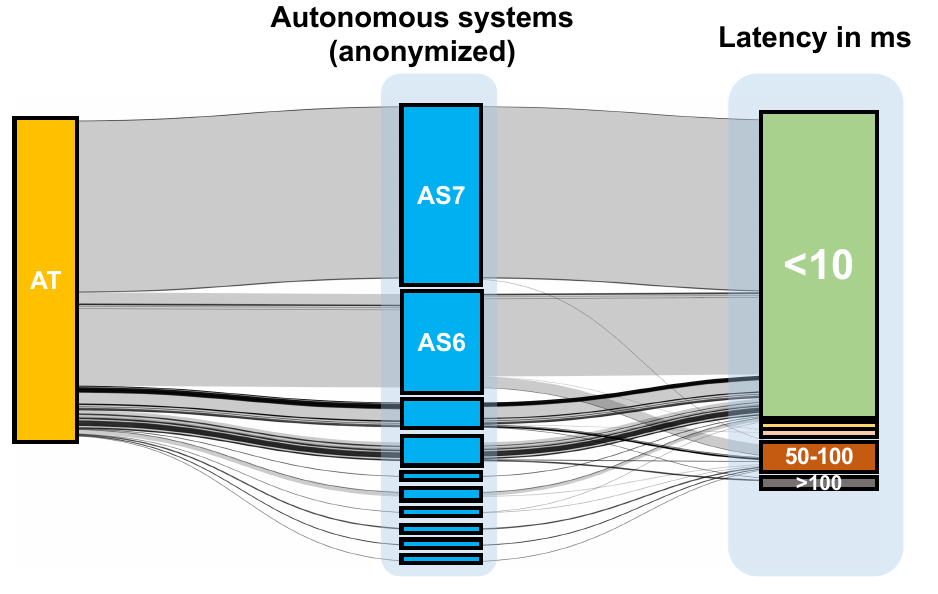}}\quad
	\label{fig:AT}
	}
	}
	\caption{Distribution of flows across autonomous systems (AS) and latency ranges for SUE, SPE, and AT states in our representative metaverse walk-through.}
	\label{fig:flowDistributionSUESPE}
\end{figure*}

When we exploring those SUEs, event content created by users and uploaded to the metaverse service domain are constantly being fetched at run-time. Therefore, we see a burst of TCP flows toward the primary domain (\ie \textit{shapevrcloud}) during SUEs shown by blue lines in Fig.~\ref{fig:flowProfile}. The respective packet rate (\ie blue region in Fig.~\ref{fig:TCPPkt}) and bandwidth usage also stay at the highest level among all states. In our walk-through, we were not always having a smooth experience during SUE states, seeing for example blank or low-resolution items for a while before it gets clearer. As depicted in Fig.~\ref{fig:flowDistributionSUE}, it is probably due to relative high latency (\ie >50ms) and limited TCP bandwidth for some primary domain flows going through AS1 that is hosting content for those items. 

Similar to MH as discussed in \S\ref{sec:MH}, there are one or two concurrent UDP flows toward time-critical activity domain of the metaverse to synchronize real-time user activities. The upstream packet rate and bandwidth stay at a constant level, while the downstream ones depend on how crowded the event is. During our gestural and vocal interactions with other users, there is no perceived feeling of lag and interruption, likely because the latency is low (\ie <10ms) and sufficient bandwidth is available for the flows hosted in AS2 as per Fig.~\ref{fig:flowDistributionSUE}.

We have also seen many decorative pictures, videos, icons, and external links embedded in each of the events. For example, in the theatre (SUE4-5), the movies being played and dynamic posters on the walls are all from external sources (\ie \textit{vimeo} and \textit{gvt2}). In the shopping mall (\ie SUE7-9), each store front is a static or dynamic picture of the brand which is linked to its official site. As a result, many flows toward different third-party service and cloud content domains are initialized during SUEs; however, the number of TCP and UDP packets to such domains is fairly low as seen in Fig.~\ref{fig:TCPPkt} and Fig.~\ref{fig:UDPPkt}. Fig.~\ref{fig:flowDistributionSUE} shows that those flows are sent to four different ASes, with some of them having latency higher than 100ms, which can lead to laggy and unresponsive experience for users in interacting with the items.

SUE is the most important state, central to the concept of the metaverse enhanced by third-parties. It is also the most demanding one among all states in terms of network resources. Sufficiently high bandwidth, low latency, and stable jitter are required for both TCP and UDP flows toward primary, time-critical activity, third-party service, and cloud content domains that all play critical roles in user-perceived experience. Moreover, the service domains involved in this state are often highly distributed among many ASes, making it challenging for network operators to optimize their routing paths to support good user experience.

\subsection{Separate Provider-Created Event (SPE)}\label{sec:SPE}
As discussed in our walk-through in \S\ref{sec:WalkExample}, apart from separate metaverse events created by users, we have also explored some events that are directly created and operated by the metaverse provider including a multi-player shooting game (SPE14) and a NASA space gallery (SPE15). Those provider-created events are often rich in content, large in space, with mechanisms more complex than the counterparts created by individual users. They are also quite attractive to users who seek casual entertainment (\eg gaming) and exploration (\eg gallery) rather than socialization (\eg meetup, conference, and ceremony) that is main focus of SUE.

The provider-created nature of SPE suggests that the respective event content could be mostly pre-downloaded as patches of the console application in user's VR headset, instead of being fetched at run-time. Thus, there is no surge of new flows to the primary domain, as shown in SPE14 and SPE15 of Fig.~\ref{fig:flowProfile}. 
We could also observe very low profiles of packet rate and bandwidth usage (such as those shown in Fig.~\ref{fig:TCPPkt}) during SPEs except their beginning periods.
Similar to MH and SUE, a user is interacting with others in real-time. Therefore, one or two UDP flows toward time-critical activity domains are always active and transmitting data at a stable rate -- a slightly higher packet rate and bandwidth usage is observed in the shooting game (\ie SPE14 of Fig.~\ref{fig:UDPPkt}) that is more demanding on real-time synchronization than other social events.

The third-party service and cloud content domains being accessed during SPEs are primarily for VR graphics such as \textit{graph-facebook-hardware}, \textit{graph.oculus}, \textit{unity3d}, and \textit{oculuscdn}. Their packet rates and bandwidth consumption are at low levels as shown in Fig.~\ref{fig:TCPPkt}. Unlike MH and SUE, there is often no or little other third-party content like advertisements being fetched during SPEs.

The user-perceived experience in SPE is not very sensitive to TCP metrics since the packet rate and bandwidth usage for VR graphics is lower than 30pps and 10Kbps. In contrast, UDP latency, bandwidth, and jitter do play important role in user-perceived experience as real-time activity synchronization among users are critical in SPEs. In addition, the optimization of routing path is not very challenging to network operators as there are only few (\ie three in our walk-through) ASes involved as shown in Fig.~\ref{fig:flowDistributionSPE}, which correspond to primary metaverse content (AS1), time-critical activity (AS2), and VR graphics (AS3).

\subsection{Asset Trading (AT)}\label{sec:AT}
Asset trading is the state where individual and commercial users could purchase and sell their digital (\eg NFT) or physical assets via third-party platforms using cryptocurrency, digital tokens, or real-world money.

In ATs, a user often stays static when browsing items on sale (typically through pop-up windows) without frequent interaction with the environment and other users. Therefore, very few packets are exchanged between a user's VR headset and the primary or time-critical activity domains, though those flows remain active as visually shown in Fig.~\ref{fig:TCPPkt} and Fig.~\ref{fig:flowProfile}.

Unlike other activity states that are dominated by primary metaverse content and time-critical activities, the majority of network traffic in ATs is for third-party service and cloud content domains. They serve the description/advertisement of items (\eg \textit{adsrvr}, \textit{pinimg}, \textit{fbcdn}, and \textit{cloudflare}) and currency platforms (\eg \textit{crypto} and \textit{opensea}). Those service domains are often located in a diversified set of ASes as visually shown in Fig.~\ref{fig:AT} -- AS7 and AS6 are for the market domain \textit{opensea} and cloud content \textit{fbcdn} respectively, while other domains belong to eight ASes, most of which have poor latency larger than 50ms.

While there are time-critical activity and primary domain flows, ATs have almost no requirement of UDP/TCP performance due to their inactivity in metaverse content/activity exchange, while TCP metrics toward third-party and cloud content domains are quite important. Given the diversity of ASes involved in this state, optimization of routing could be complex to achieve.

\begin{wrapfigure}{L}{0.45\textwidth}
	{\includegraphics[width=0.45\textwidth]{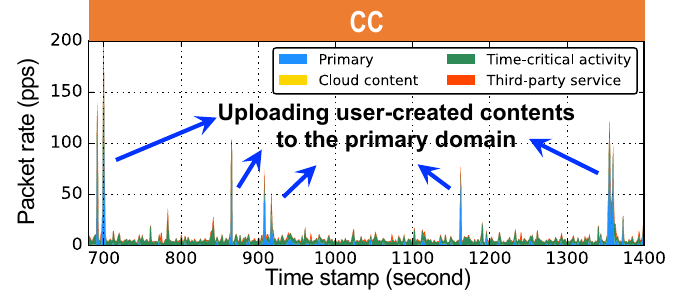}}
	\vspace{-6.5mm}
	\caption{Packet rates during content creation (CC) activity in a Rec Room session.}
	\label{fig:recRoomCC}
\end{wrapfigure}

\subsection{Content Creation (CC)}\label{sec:CC}
Content creation by users in Multiverse is done outside of the metaverse, and thus there is no CC state during our representative walk-through in \S\ref{sec:WalkExample}. The Rec Room metaverse allows content creation to be done within the application, and we conducted a session to illustrate the network behavior. The traffic profile of the CC state is shown in Fig.~\ref{fig:recRoomCC}, where the user is creating an outdoor party using available building blocks. In short, the primary domain during CC occasionally has a quite spiky pattern for its upstream packet rate to update the content created by a user to the metaverse primary domain \textit{rec}. 

The time-critical domain flows are active during CC, but with much less packet rate and bandwidth consumption (possibly for keep-alive messages) compared with MH, SUE, and SPE, as there is no interaction with other users when creating content. We could not observe any cloud content domain in this state as it does not load any content provided by other users, and there is a small amount of traffic exchanged with third-party service domain \textit{oculus} for the VR platform.

In terms of network requirements, CC requires sufficient TCP bandwidth to ensure timely content upload from user device to the primary metaverse domain,which could go up to 15Mbps in our representative Rec Room session.

\subsection{Summary}\label{sec:DevelopmentSummary}
The key takeaways from our analysis of network traffic in metaverse VR applications are as follows. 
Firstly, during a typical session, there are often over several tens of service domains being accessed. These domains can be categorized into primary, time-critical activity, cloud content, and third-party domains based on their purposes and ownership. Due to the complex composition of these domains, their hosting servers are often highly distributed across ASes reached via diverse routing paths. However, some of these paths may not be well-optimized for user experience, such as latency. 
Secondly, among the six distinct metaverse user activity states, HS (Home Space) and SUE (Separate User-created Event) are indispensable in all metaverse VR applications. MH (Main Hub) is a common state that helps users transition between events. On the other hand, AT (Asset Trading), CC (Content Creation), and SPE (Separate Provider-created Event) are currently only available in a few (one or two) applications. 
Thirdly, flows from the four types of service domains exhibit distinguishable patterns depending on the user activity states. For example, primary flows responsible for fetching/uploading metaverse event content are highly active with significant volumetric profiles during SUEs, the initial phase of MHs and SPEs, as well as periodically in CCs. On the other hand, time-critical activity flows that consistently update user motions remain active in MHs, SUEs, and SPEs but are mostly idle otherwise. These unique traffic patterns can be leveraged to classify metaverse user activity states, as discussed next.

\section{\textit{MetaVRadar:} Detecting and Classifying User Activity State}\label{sec:characteristics}
In this section, we discuss how the above-studied insights inspire our design of \textit{MetaVRadar} that detects metaverse VR sessions and classifies user activity states. Our approach is designed to withstand not just HTTPS packet content encryption, but also encryption of the SNI carried in TLS client hello packets, as envisaged by HTTP/3 with encrypted client hello (ECH). We additionally eliminate all reliance on DNS, not just due to encryption, but also because many UDP flows are not preceded by DNS lookups. Our method instead relies on extracting a wide range of flow-level attributes that are encryption agnostic. To detect metaverse activity (\S\ref{sec:trafficSignature}) we build signatures based on packet payload size sequences of primary and time-critical activity domain flows. In \S\ref{sec:networkAttributes} we develop volumetric attributes pertaining to the network behavioral profiles of the various activity states of the metaverse, which leads us to design a three-stage detection and classification methodology in \S\ref{sec:methodology}.

\subsection{Packet Payload Size Sequence Signatures of Metaverse Flows}\label{sec:trafficSignature}
As discussed in the prior section, flows toward primary and time-critical activity domains that are directly operated for each metaverse hold unique patterns in their first several packet payload sizes, which could be used to accurately detect those metaverse flows and identify their domain types. As for primary domain flows, the packet size-based signatures are robust to SNI encryption that could make their domain names and service prefixes not observable by network operators.

\begin{table*}[t!]
	\centering
	\caption{Prefix sequence and packet payload size signatures of flows toward primary domains of each metaverse application, which contains \textbf{\color{purple}TLS client hello}, crypto key exchange (CKE), crypto key selection (CCS), encrypted handshake message (EHM) and \textbf{\color{blue}first application data}.}
	\small
	\begin{tabular}{|p{1.5cm}|p{1.6cm}|p{9.3cm}|}
		\hline
		\rowcolor[rgb]{ .906,  .902,  .902} 	\textbf{Metaverse}     & \textbf{Primary domain}  & \textbf{Service prefix sequence of primary domain flows and their upstream packet payload size sequence signatures} \\ \hline
		Multiverse   &            \textit{shapevrcloud}  & \textit{prod}: [\textbf{\color{purple}414},75,6,45,\textbf{\color{blue}338}] $\rightarrow$\textit{prod}: [\textbf{\color{purple}414},75,6,45,\textbf{\color{blue}591}] $\rightarrow$ .. $\rightarrow$ \textit{prodblobs}: [\textbf{\color{purple}419},75,6,45,\textbf{\color{blue}284}] \\ \hline
		VRChat       &             \textit{vrchat} & \textit{api}: [\textbf{\color{purple}409},75,6,45,\textbf{\color{blue}235}] $\rightarrow$ .. $\rightarrow$ \textit{pipeline}: [\textbf{\color{purple}244},134,\textbf{\color{blue}490}]  $\rightarrow$ \textit{assets}: [\textbf{\color{purple}410},75,6,45,\textbf{\color{blue}305}]    \\ \hline
		Rec Room     &       \textit{rec}        &  \textit{api}: [\textbf{\color{purple}148},75,51,\textbf{\color{blue}204}]   $\rightarrow$ \textit{api}: [\textbf{\color{purple}148},75,51,\textbf{\color{blue}205}]   $\rightarrow$ .. $\rightarrow$ \textit{api}: [\textbf{\color{purple}148},75,51,\textbf{\color{blue}253}]   $\rightarrow$ \textit{auth}: [\textbf{\color{purple}149},75,51,\textbf{\color{blue}216}]    \\ \hline
		AltSpaceVR &    \textit{altvr}    & \textit{config}: [\textbf{\color{purple}409},75,6,45,\textbf{\color{blue}269}]  $\rightarrow$ \textit{cdn-content-ingress}: [\textbf{\color{purple}422},107,6,45,\textbf{\color{blue}276}]  $\rightarrow$ \textit{account}: [\textbf{\color{purple}410},107,6,45,\textbf{\color{blue}239}]  \\ \hline
	\end{tabular}
	\label{tab:sampleByteSignatures}
\end{table*}

\subsubsection{Primary domain flows}\label{sec:byteSequencePrimaryDomain}
At the beginning of a metaverse VR session, the user device (\ie VR headset) initializes a series of TLS-encrypted TCP flows to exchange specific administrative information (\ie authentication and user activity tracking) with the primary domain. The first several upstream packets in those flows are identifiable by their packet payload sizes excluding Ethernet, IP and transport-layer headers. 
Specifically, those flows are started by TCP and TLS handshakes, followed by the actual application data specific to the request services.
TCP handshakes consist of standard SYN and ACK packets with no payload content (\ie zero byte). In a TLS handshake, the user sends two to five upstream packets specific to the requested service, including client hello, crypto key exchange, crypto key selection and encrypted handshake -- the latter three messages could be delivered by either one or separate packets. After the handshake processes, the first content packet carrying application-specific data also has a certain payload size, which is hardly surprising as it often serves as a fixed starting message of the requested service.
Given the fixed content to be exchanged in the first several upstream packets for TLS handshake of the metaverse administrative flows, as will be evaluated in \S\ref{sec:evalDetectionAccuracy}, the payload size sequence could be used as a reliable signature to infer the requested service domain when SNI becomes not observable.

\textbf{Extracting flow signatures:} 
As discussed in \S\ref{sec:HS}, at the start of a metaverse VR session, there is a series of primary domain prefixes that are accessed through multiple flows from the user, each having an important role (\eg authentication and critical application content) for the session. We developed an automatic training process (using Golang and Python) that takes ground-truth traffic traces files (\ie PCAPs) and their primary domain names (\eg \textit{altvr} for AltSpaceVR) as inputs to extract packet size sequence signatures (containing packet payload sizes of TLS handshakes and first application data) of the primary domain flows with their respective prefix names.

\textbf{Representative packet payload size sequences:} 
Obtained from our training process, some representative per-flow signatures toward primary domains of the three studied metaverses are given in Table~\ref{tab:sampleByteSignatures} as concrete examples. The upstream packet payload sizes of each primary domain flow during the first HS state are listed in separate brackets each labeled by their service prefixes in the order of appearance. The payload sizes of TLS client hello packets, other TLS handshake packets (\ie CKE, CCS and EHM), and first application data packets are color-coded as purple, black and blue in Table~\ref{tab:sampleByteSignatures}, respectively. Taking the Multiverse as an example, in our training trace files, a user always sends six flows toward the service prefix \textit{prod} followed by one flow towards \textit{prodblobs}. The first flow has 414 bytes of payload in its TLS client hello packet, 75 bytes of payload in the CKE packet, 6 bytes in the CCS packet, 45 bytes in the EHM packet, and 338 bytes in the first application data packet. We note that not all primary domain flows have their TLS CKE, CCS, and EHM in separate packets. As shown in Table~\ref{tab:sampleByteSignatures}, the flow towards the ``\textit{pipeline}'' prefix of VRChat has all three messages in one packet with a payload size of 134 bytes, whereas the primary flows for Rec Room have their CCS and EHM messages in one packet with a payload size as 51 bytes. 

Later in \S\ref{sec:methodology}, we will discuss how the upstream packet payload size signatures are used to detect active primary domain flows toward different service prefixes. The sequence of service prefixes accessed by the detected flows is a robust indicator of active metaverse sessions. 

\subsubsection{Time-critical activity domain flows}\label{sec:byteSequenceTimeCriticalActivityDomain}
UDP flows toward time-critical activity domains are created during states when the user is synchronizing her/his positions and gestures with others.
Such flows are sent to certain port numbers and with unique payload size sequences of their first several (\eg four to seven) upstream packets. These packets carry initial messages specific to the requested services, similar in spirit to multiplayer gaming applications \cite{SCMadanapalliPAM2022}. We therefore extract the per-flow upstream packet payload size sequences for traffic destined to the various time-critical activity domains UDP ports (\eg \textit{5055}, \textit{5056}, and \textit{5058} used for Multiverse), and train our machines using a similar method as described in \S\ref{sec:byteSequencePrimaryDomain}. Example packet size sequence signatures of flows toward those UDP port numbers are shown in Appendix~\ref{sec:timecriticalDomainByteSignature}. 

In \S\ref{sec:methodology}, we will discuss how our proposed methodology uses the identified primary and time-critical activity domain flows to detect active metaverse sessions and track their activity states.

\begin{figure*}[t!]
	\begin{center}
		\mbox{
			\hspace{-3mm}
			\subfigure[Systematic definition of attributes.]{
				{\includegraphics[width=0.48\textwidth]{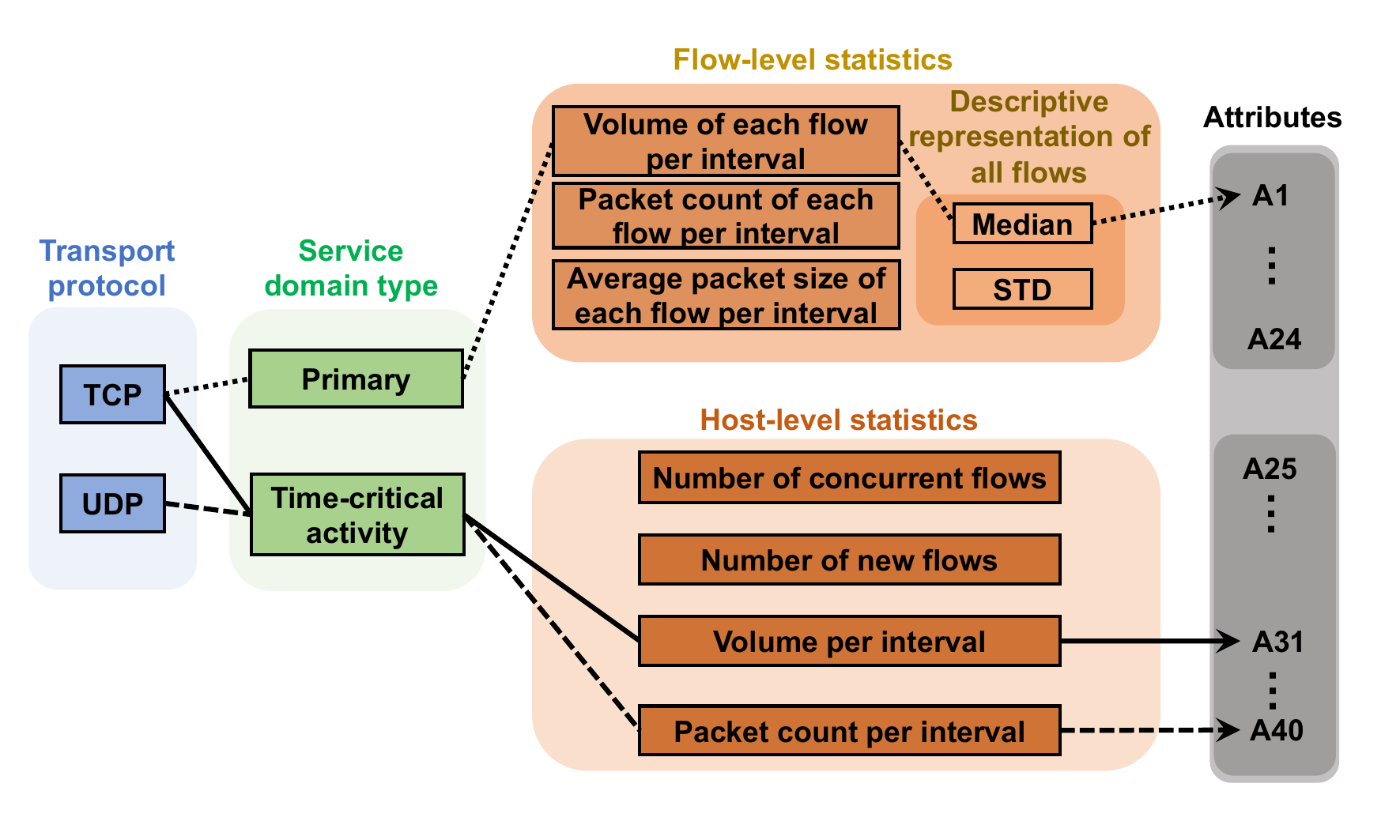}}\quad
				\label{fig:attributeDefinition}
			}
			\hspace{-3mm}	
			\subfigure[Representative values of attributes for Multiverse.]{
				{\includegraphics[width=0.48\textwidth]{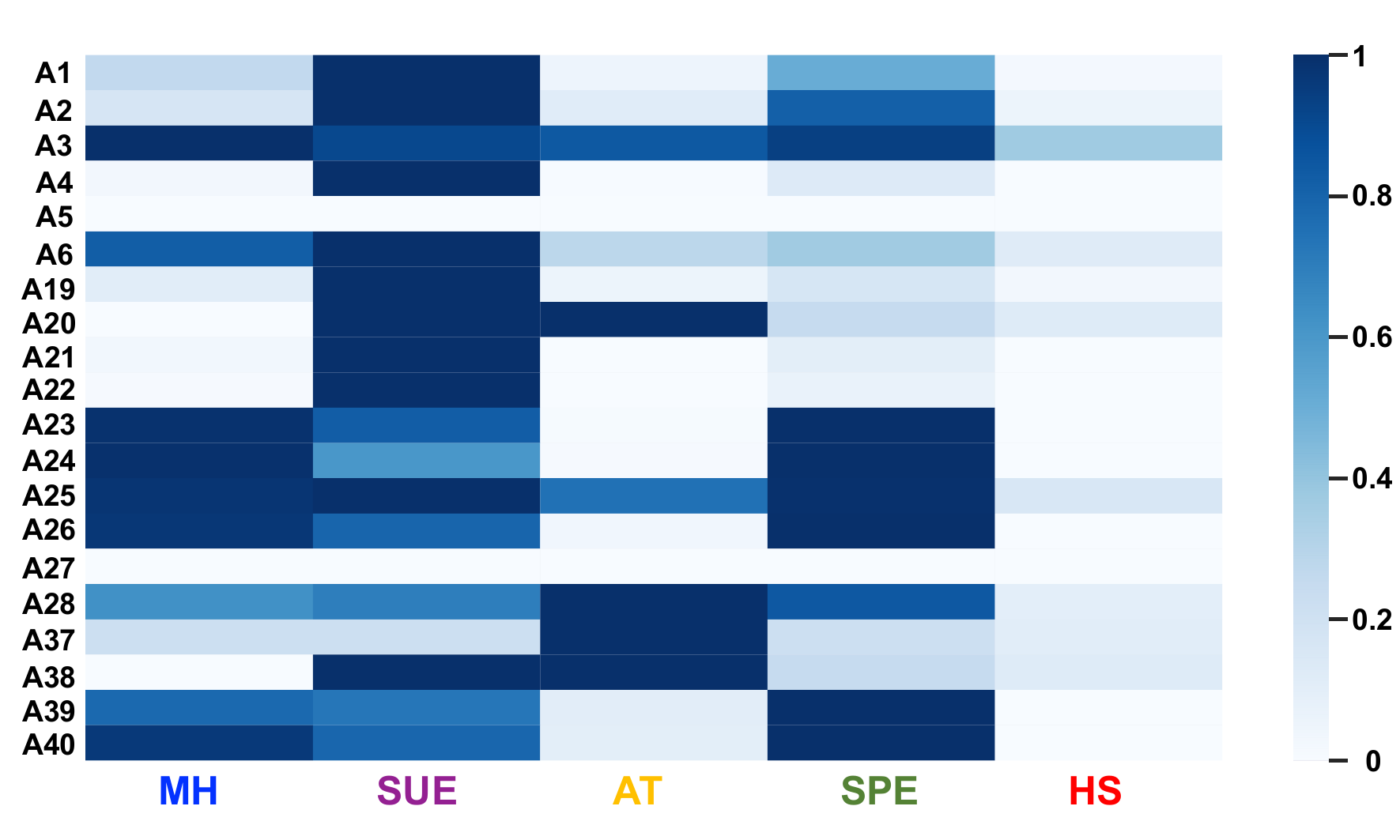}}\quad
				\label{fig:attributeValue}
			}
		}
		\caption{Systematic definition of network traffic attributes of metaverse user activity states and their representative normalized values for Multiverse. A full list of attributes is provided in Appendix \S\ref{sec:AppendixAttributeList}.}
		\label{fig:TrafficAttribute}
	\end{center}
\end{figure*}

\subsection{Volumetric Attributes of Metaverse User Activity States}\label{sec:networkAttributes}
Driven by the discussion in \S\ref{sec:HS} -- \S\ref{sec:CC}, different states of user activity exhibit unique volumetric patterns that structured attributes could capture. We systematically define forty attributes from packet and flow statistics of metaverse sessions that could be visually shown in Fig.~\ref{fig:attributeDefinition}. 

Our attributes cover transport-layer protocols (\ie TCP or UDP), service domain types that are directly operated for a metaverse (\ie primary and time-critical activity), and various measurement metrics of packets and flows. Those metrics are either directly maintained for each session (\ie host-level statistics in Fig.~\ref{fig:attributeDefinition}); or for all user flows that are represented by their median and standard deviation values. 
For example, the attribute A1 is defined by the top dotted line in Fig.~\ref{fig:attributeDefinition}, which in defined as the median volume of all TCP flows toward primary domains during a monitored interval. Similarly, the last attribute A40 is defined by the bottom dashed line as UDP packet count toward time-critical activity domain from a metaverse session during a monitored interval (\eg 10 seconds in our implementation). 
A full list of attributes is provided in Appendix~\S\ref{sec:AppendixAttributeList}.

The attributes collectively cover distinguishable volumetric characteristics of a metaverse session in different activity state. We shown a representative set of attribute values under different states as computed from our Multiverse traffic traces in Fig.~\ref{fig:attributeValue} (these values have been normalized to their maximum purely for visualization purpose). Each activity state has their own attribute value ranges as color-coded in Fig.~\ref{fig:attributeValue}, except twenty attributes that are excluded in the heatmap as they remain zeros for Multiverse. It is because of that the zero-valued twenty attributes are for TCP of primary domain and UDP of time-critical activity domain, which do not exist in the four studied metaverses. However, they are reserved to capture future variations such as QUIC over UDP for primary metaverse content.

\subsection{Methodology of Metaverse Session Detection \& User Activity State Classification}\label{sec:methodology}
\begin{figure*}[t!]
	\includegraphics[width=\textwidth]{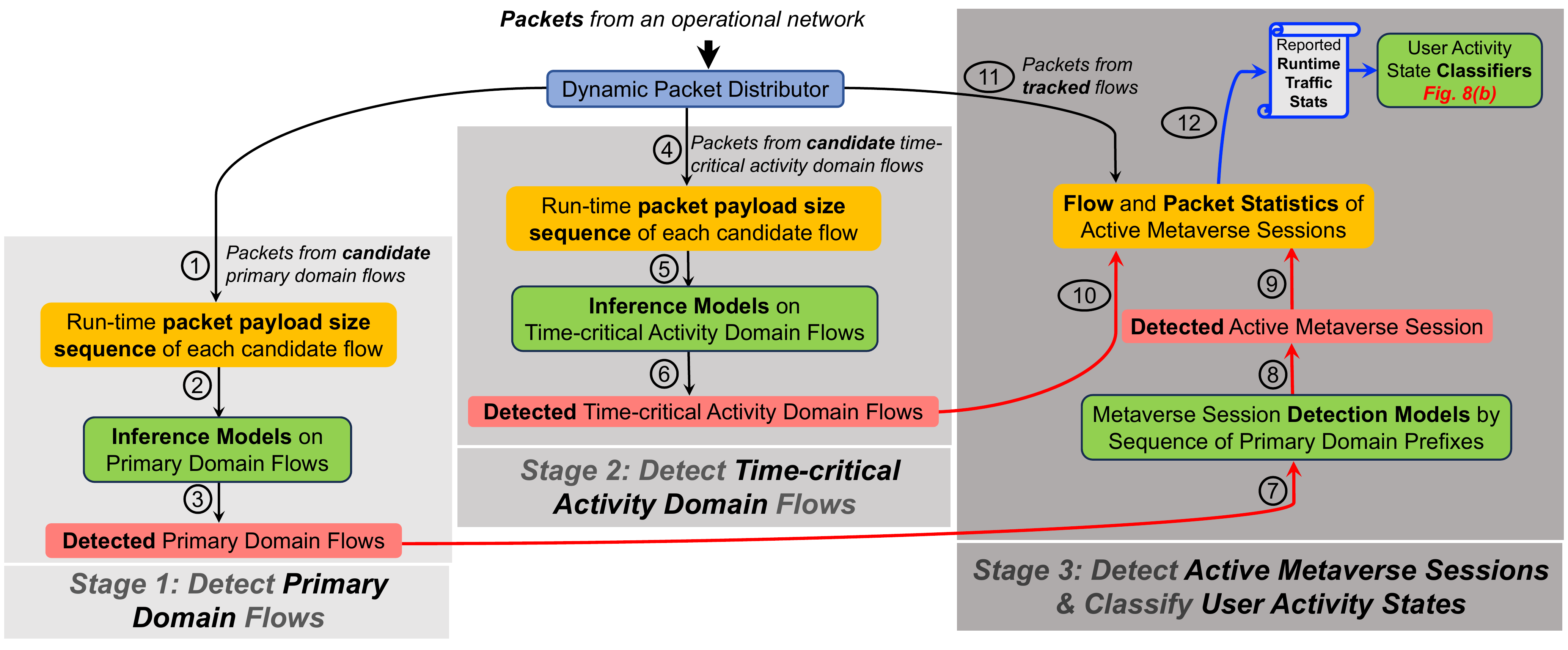}
		\vspace{-4mm}
	\caption{Our three-stage methodology for the real-time detection and classification of user activity states.}
	\label{fig:DetectionPipeline}
\end{figure*}

Fig.~\ref{fig:DetectionPipeline} shows our proposed method \textit{MetaVRadar} that receives a packet stream (\eg from a tap or mirror) from an operational network to detect and classify metaverse user activities through three stages, each of which has a unique objective and scope of processed packet streams, thus, is equipped with its own runtime data structure (yellow boxes) and inference models (green boxes).

At a high level, the first stage identifies flows to primary domain and their service prefixes, while the second stage identifies flows toward time-critical activity domains. The third stage uses the outputs of the first two stages to identify metaverse sessions and classify user activity states, as described in more detail below.

\subsubsection{First Stage: Detecting Primary Domain Flows}
The first stage detects metaverse flows toward primary domains.
In the step \textcircled{1}, this stage receives upstream packets from an operational network (\eg through a packet mirroring configuration on the edge router) with filtering conditions as TCP flows to destination port 443, the potential candidates of primary flows.

In the step \textcircled{2}, the received packets will be sorted into a runtime data structure (\eg a hash table in our implementation) with two key layers indexed by user IP and flow metadata (5-tuples), respectively. The value field of the data structure is the sequence of upstream packet sizes of this flow.
At runtime, the updated packet payload size sequence of a flow will be checked against a pre-trained inference model -- a collection of size sequences covering TLS handshakes and first application data, as discussed in \S\ref{sec:byteSequencePrimaryDomain}. Each packet size sequence in our inference model is mapped to a primary domain prefix and name, \eg [\textbf{\color{purple}409},75,6,45,\textbf{\color{blue}269}] is mapped to the \textit{config} prefix of the domain name \textit{altvr} as given in Table~\ref{tab:sampleByteSignatures}, wherein the first large size (409 bytes) is for TLS client hello, the three following small sizes are for CKE, CCS and EHM, and the fifth size (269 bytes) is for the first application data. The perfect match of a flow packet size sequence will trigger a successful detection. As the step \textcircled{3} in Fig.~\ref{fig:DetectionPipeline}, metadata of the detected flows including user IP, service prefixes and flow 5-tuples will be sent to the third stage.

\subsubsection{Second Stage: Detecting Time-Critical Activity Domain Flows}
The second stage detects metaverse flows toward time-critical activity domains. In step \textcircled{4} , all upstream UDP packets with their destination service port numbers matched in a pre-listed range (discussed in \S\ref{sec:byteSequenceTimeCriticalActivityDomain}) obtained from ground-truth training traffic traces will be received by this stage. Similar to the first stage, in step \textcircled{5}, a run-time data structure containing two key layers of user IP and flow 5-tuples tracks the size sequences of UDP packets in each flow and reports the updated size sequence to a pre-trained inference model for detection. In step \textcircled{6}, the metadata of detected flows (\ie user IP, metaverse name, flow 5-tuples) that have matched ports and size sequences will be sent to the third stage. 

\subsubsection{Third Stage: Detecting Metaverse Session and Classifying User Activity State}\label{sec:thirdStage}
The third stage detects and classifies active metaverse VR sessions to give operators visibility in their networks. The detection is based on tracking service prefixes of active primary flows, and the classification is achieved by analyzing volumetric attributes of active metaverse VR sessions through a stateful approach using machine-learning models.

\begin{wrapfigure}{L}{0.45\textwidth}
	{\includegraphics[width=0.45\textwidth]{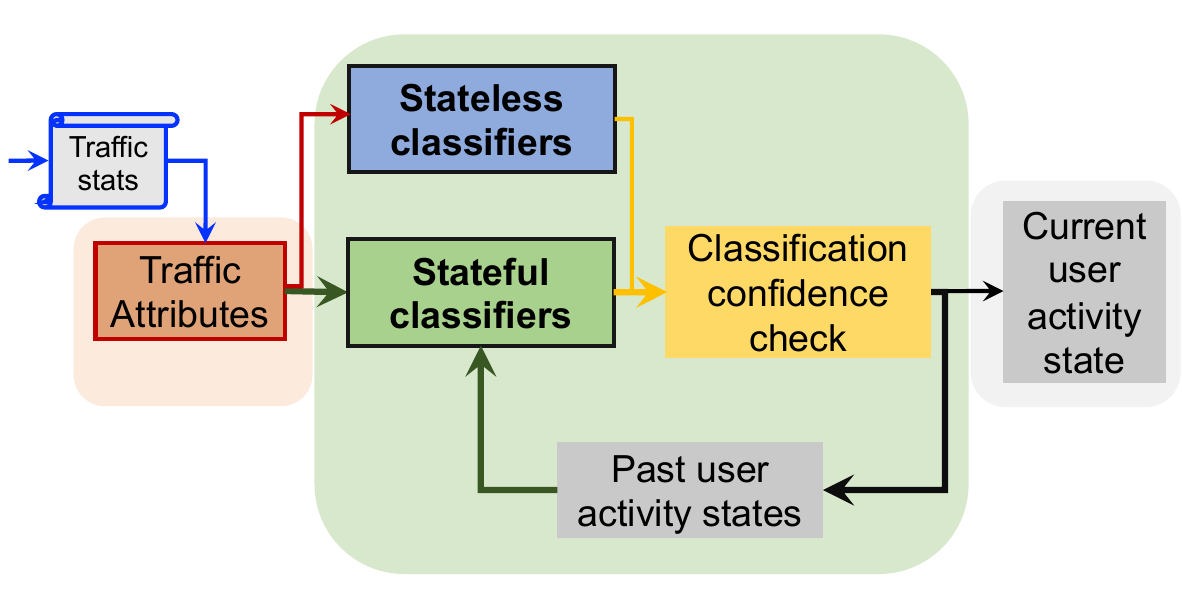}}
	\vspace{-4mm}
	\caption{Classification methodology of user activity states via stateful approach with ML classifiers.}
	\label{fig:classification}
\end{wrapfigure}

\textbf{Detection via prefix sequence of primary domain flows:} 
As discussed in \S\ref{sec:HS}, primary flows always start in the beginning of each metaverse session for authentication and initialization. Thus, in our detection models highlighted in the green box in Fig.~\ref{fig:DetectionPipeline}, we track the primary domain prefixes of metaverses that have active flows from each session as a detection criterion. When all primary domain prefixes that should appear in the initial HS states (\ie shown in Table~\ref{tab:sampleByteSignatures}) have been marked as active from the detection results of stage 1 (step \textcircled{7}), a corresponding entries will be created in the run-time data structure (step \textcircled{8} and \textcircled{9}) to track the network stats of this detected metaverse session.
A detected time-critical activity domain flow (from stage 2) of a tracked metaverse session will add its corresponding entry in the run-time data structure (step \textcircled{10}).

In step \textcircled{11}, volumetric stats of the tracked flows including their packet counts and volumes over a certain interval (\eg 10 second in our implementation) are updated according to the arrived packets, which are periodically sent to the classification module to determine user activity states (step \textcircled{12}).
We emphasize that the detection of metaverse sessions in our approach solely relies on the detection results of primary domain flows in stage 1. The time-critical activity domain flows detected in stage 2 are utilized for user activity classification in stage 3.

\textbf{Classification via volumetric attributes:}
Using the volumetric attributes (as described in \S\ref{sec:networkAttributes}), we have developed a stateful approach for user activity state classification. This classification process employs machine learning models implemented using the random forest algorithm. We opted for this stateful design due to the inadequate performance of our initial classification attempts using stateless approaches. The details of our model training, evaluation and selection process are described in Appendix \S\ref{sec:AppendixTrainingDetails}.
The key reason for adopting the stateful approach is that each state's unique traffic patterns may not always be consistent during all intervals. For example, MH and SPE have very spiky patterns in the use of primary domains during their beginning intervals; and CC occasionally exhibits patterns of content uploading that may not always be captured in each monitored interval.

Now we discuss the core logic of our stateful classification highlighted by the green region in Fig.~\ref{fig:classification}.
During operation, the ML classifiers infer user activity states by considering both the traffic attributes of the current interval and the past \textit{N} states (where \textit{N=5} is chosen from a range of 1 to 10 after the training process). We note that the past states refer to the classification results from previous inference intervals. However, if there are insufficient past states or the confidence level of a result from the stateful classifiers is below an administratively defined threshold \textit{T} (where \textit{T=85\%} is selected after training), the user activity state of the current interval will be determined by the stateless ML classifiers (represented by the blue module in Fig.~\ref{fig:classification}) based on traffic attributes.

\section{Evaluation And Deployment Insights}\label{sec:detectMetaApp}
We implement \textit{MetaVRadar} as a proof-of-concept prototype leveraging virtual network functions built on top of the DPDK framework. It processes real-time traffic streams from our university campus mirrored to two 10 Gbps network interfaces on a commodity server in our lab, which could be easily scaled up for a high-throughput network by adding more parallel computes. Detection signatures and classification models loaded on the prototype are generated from our training process, which could be easily updated by new and comprehensive datasets. Now we demonstrate its efficacy in detecting and classifying ground-truth metaverse user activities from our lab environment (\S\ref{sec:labEval}) as well as its practicality to be deployed in a large network (\S\ref{sec:uniEval}).

\subsection{Evaluation in the Lab Environment}\label{sec:labEval}
We evaluate the accuracy and processing overhead of our prototype in detecting and classifying ground-truth user activities in four metaverse VR applications under our lab network environment. 

\begin{table*}[!t]
	\caption{Detection accuracy (in the format of ``true positive | false positive'') of our prototype for metaverse sessions, primary domain flows and time-critical activity domain flows. The accuracy for flows are reported in four categories including `All', `Long' (\ie more than 30 seconds), `Med.' (\ie 10 to 30 seconds) and `Short' (\ie less than 10 seconds). The value `0' and `0\%' in the table indicate absolute zero and a small rounded value of zero, respectively. }
	\small
	\begin{tabular}{|p{1.37cm}|p{1cm}|llll|llll|}
		\hline
		\multirow{2}{*}{\textbf{Metaverse}} & \multirow{2}{*}{\textbf{Session}}  & \multicolumn{4}{c|}{\textbf{Primary domain flows}} & \multicolumn{4}{c|}{\textbf{Time-critical actv. domain flows}}      \\ \cline{3-10}
	&                     & \multicolumn{1}{l|}{\textbf{All}} & \multicolumn{1}{l|}{\textbf{Long}} & \multicolumn{1}{l|}{\textbf{Med.}} & \textbf{Short} & \multicolumn{1}{l|}{\textbf{All}} & \multicolumn{1}{l|}{\textbf{Long}} & \multicolumn{1}{l|}{\textbf{Med.}} & \textbf{Short} \\ \hline
		Multiverse   &          {100\%|0}    & \multicolumn{1}{l|}{99.6\%|{\color{purple}0\%}}    & \multicolumn{1}{l|}{99.7\%|{\color{purple}0\%}}     & \multicolumn{1}{l|}{99.7\%|0}     &  99.8\%|0     & \multicolumn{1}{l|}{98.2\%|0}    & \multicolumn{1}{l|}{100\%|0}     & \multicolumn{1}{l|}{100\%|0}     &    {{\color{red}67.4\%}|0}   \\ \hline
		VRChat            &         {100\%|0}      & \multicolumn{1}{l|}{90.8\%|0}    & \multicolumn{1}{l|}{{\color{red}84.4\%}|0}     & \multicolumn{1}{l|}{96.1\%|0}     & 97.5\%|0      & \multicolumn{1}{l|}{99.6\%|0}    & \multicolumn{1}{l|}{99.1\%|0}     & \multicolumn{1}{l|}{100\%|0}     &    100\%|0   \\ \hline
		RecRoom           &          {100\%|0}      & \multicolumn{1}{l|}{94.0\%|{\color{purple}0\%}}    & \multicolumn{1}{l|}{89.3\%|0}     & \multicolumn{1}{l|}{95.6\%|0}     &  --|{\color{purple}0\%}     & \multicolumn{1}{l|}{97.1\%|0}    & \multicolumn{1}{l|}{94.8\%|0}     & \multicolumn{1}{l|}{99.5\%|0}     &   98.5\%|0    \\ \hline
		AltSpaceVR        &        {100\%|0}       & \multicolumn{1}{l|}{94.9\%|0}    & \multicolumn{1}{l|}{87.2\%|0}     & \multicolumn{1}{l|}{90.7\%|0}     &  99.1\%|0     & \multicolumn{1}{l|}{96.5\%|0}    & \multicolumn{1}{l|}{100\%|0}     & \multicolumn{1}{l|}{{\color{red}81.0\%}|0}     &   100\%|0    \\ \hline
	\end{tabular}
	\label{tab:detectionAccuracy}
\end{table*}

\subsubsection{Detection Accuracy}\label{sec:evalDetectionAccuracy}
During our lab evaluation, we started 20 sessions of each metaverse studied in this paper, as well as four types of other networked applications for comparative testing, including PC visits to metaverse websites, VR games in the Oculus platform, online PC games, and downloads of metaverse VR console. The accuracies of our prototype in detecting primary domain flows (stage 1 in \textit{MetaVRadar}), time-critical activity domain flows (stage 2), and active metaverse sessions (stage 3) are reported in Table~\ref{tab:detectionAccuracy}. Each cell in the table represents a combination of true positive (TP) and false positive (FP) rates in the format of $TP|FP$.

First, since our method detects metaverse session with rigorous matching criteria of domain prefix sequence from primary flows, all metaverse VR sessions can be correctly detected with $100\%$ TP rates. No other application session is misidentified as metaverse, resulting in $100\%|0$ in all four cells under the ``Session'' column.
We also note that our prototype also has $100\%$ temporal coverage of each metaverse VR session by collectively considering primary and time-critical activity flows. 

We now look at the detection accuracy of primary and time-critical domain flows, which are reported under four categories of duration, including all, long (\ie $>$30s), medium (\ie from 10s to 30s), and short (\ie $<$10s). In general, flows of both domain types for all four metaverses have been accurately detected, mostly with >95\% TP and absolute 0 FP. We also note that five long and seven short non-metaverse flows are misidentified as primary flows of Multiverse and Rec Room, respectively. Given the large number (\ie more than 1M) of non-Meta flows in our evaluation dataset, the two FPs are labeled as $0\%$ in Table~\ref{tab:detectionAccuracy}. 
In addition, long primary flows of VRChat, short time-critical activity flows of Multiverse, and medium time-critical activity flows of AltSpaceVR are detected with their accuracies lower than 85\%, as marked by red color. It might be due to the limitation of our training dataset that does not provide a very comprehensive coverage of all possible flow profiles. Such short flows are often for ad-hoc items instead of major event content. Therefore, it is important to note that missing such non-critical flows does not impact the detection of active metaverse sessions.

\begin{table}[!t]
	\caption{Classification accuracy (\ie ``true positive | false positive'') of user activity states using our \textbf{ML-based stateful classification approach} on the four studied metaverse VR applications.}
	\label{tab:classificationAccuracyStateful}
	\small
	\begin{tabular}{|l|l|l|l|l|l|l|}
		\hline
		\rowcolor[rgb]{ .906,  .902,  .902} \textbf{Metaverse} & \textbf{\color{red}HS}        & \textbf{\color{blue}MH} & \textbf{\color{purple}SUE} & \textbf{\color{olive}SPE} & \textbf{\color{orange}CC} & \textbf{\color{brown}AT} \\ \hline
		Multiverse & 100\%|0.5\% & 94\%|1.3\%   &   97\%|0.7\%  &  95\%|1.5\%   & -- & 98\%|0.0\%   \\ \hline
		VRChat       &     100\%|1.0\%        & --   &  99\%|0\%   & --  & -- & -- \\ \hline
		Rec Room     &        93\%| 1.2\%    & 99\%|0\%   &  98\%|2.0\%   & --    &  97\%|1.0\%  & 92\%|1.0\% \\ \hline
		AltSpaceVR   &  100\%|0\%  & 96\%|1.8\%   &   95\%|1.3\%  & --  & -- & -- \\ \hline
	\end{tabular}
\end{table}

\subsubsection{Classification Accuracy} 
Initially, we developed stateless classifiers solely based on network attributes extracted per inference interval. However, these classifiers did not yield satisfactory accuracy, particularly in the case of CC, MH, and SPE, where consistent network profiles were not always present. In Appendix \S\ref{sec:AppendixClassification}, we provide a baseline evaluation that includes the details of the stateless method.
To address this limitation, we subsequently developed a stateful approach (\S\ref{sec:thirdStage}) in our prototype, which demonstrated high accuracy in user activity state classification across all studied metaverses. Table~\ref{tab:classificationAccuracyStateful} presents the accuracy by this stateful approach. Notably, it achieved a true-positive rate of over 95\% or close to it for all user activity states, while maintaining a false-positive rate of less than 2\%.

\subsubsection{Real-time Processing Overhead}\label{sec:realTimeOverHead}
We measure the processing time of individual modules (\ie session detection, runtime traffic statistics, and user activity classification) to process an active metaverse session. The benchmarks are conducted on our commodity server with the specifications outlined in \S\ref{sec:labSetup}. The results demonstrate that the three modules consume less than 3ms, 5-8ms, and 18-28ms per ten seconds (with standard deviations of 0.02ms, 1.25ms, and 3.42ms) for each inference cycle on a single computing thread, respectively. Therefore, our prototype can achieve real-time inferencing for over 12K active metaverse sessions on our benchmarking server with dual processor 16-core CPU. In commercial settings, a network operator may choose to use parallel servers and reduce inference frequency to further scale it up.

\subsection{Deployment Insights in a Campus Network}\label{sec:uniEval}
After evaluating the performance of \textit{MetaVRadar} in our lab environment, we demonstrate its capability in an operational network (\ie our university campus) with millions of concurrent flows and tens of Gbps traffic throughput in real-time. 
Our prototype currently focuses on flows toward primary and time-critical domains that are directly operated for metaverse VR applications and could be extended to third-party service and cloud content domains visited by active metaverse sessions. However, tracking general-purpose domain usage could be inaccurate for metaverse sessions behind NAT.
\textit{MetaVRadar} processes packets mirrored from our campus network and provides operators visibility into emerging metaverse VR (such as in application popularity, bandwidth usage, user habit, service domain distribution and latency) for optimization of their network management.

 \begin{figure*}[t!]
	\begin{center}
		\mbox{
			\hspace{-5mm}
			\subfigure[Tracked flow (ground-truth sessions).]{
				{\includegraphics[width=0.46\textwidth]{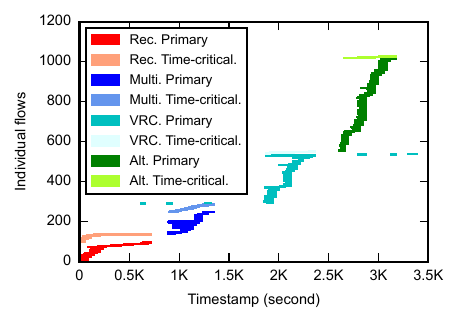}}\quad
				\label{fig:fieldFlow-6Sep}
			}
			\subfigure[Tracked packet rate (ground-truth sessions).]{
				{\includegraphics[width=0.44\textwidth]{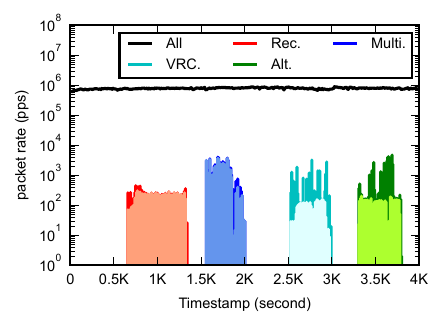}}\quad
				\label{fig:fieldPkt-6Sep}
			}
		}
		\mbox{
			\hspace{-5mm}
			\subfigure[Tracked packet rate (the University dormitory).]{
				{\includegraphics[width=0.5\textwidth]{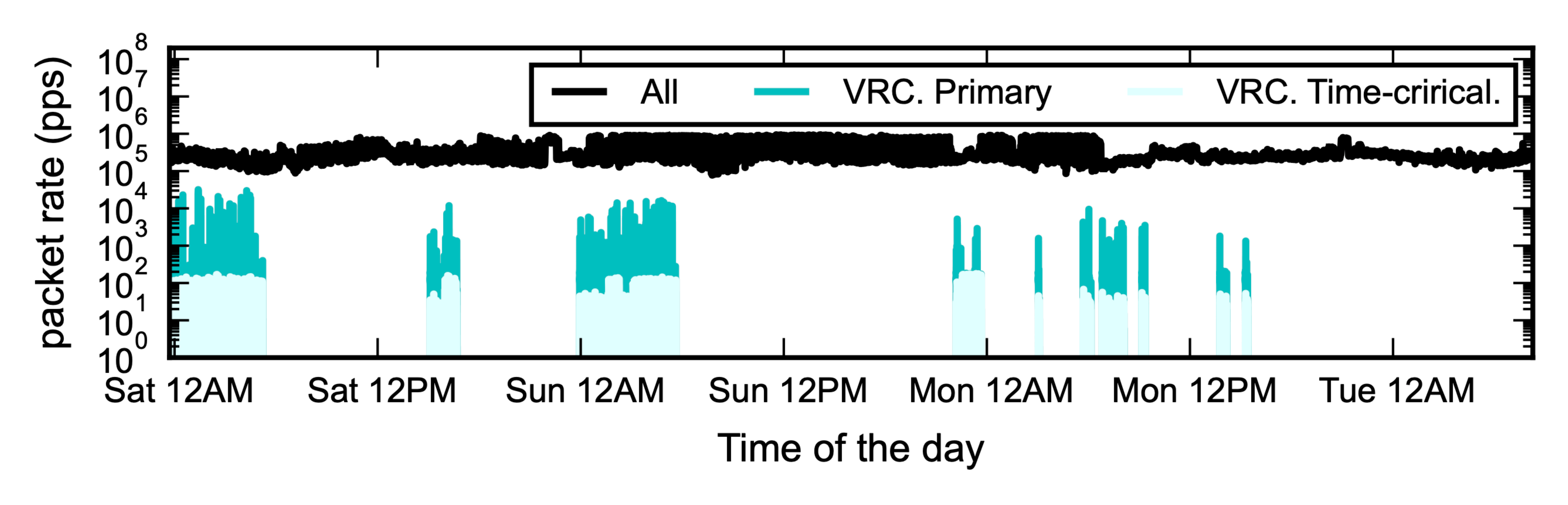}}
				\label{fig:fieldPkt-dorm}
			}
			\subfigure[Volume consumed.]{
				{\includegraphics[width=0.18\textwidth]{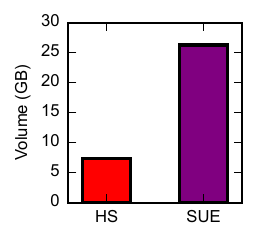}}
				\label{fig:fieldPkt-dorm-profile-1}
			}
			\hspace{-1mm}	
			\subfigure[Time spent.]{
				{\includegraphics[width=0.18\textwidth]{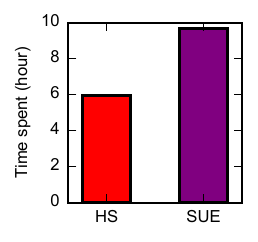}}
				\label{fig:fieldPkt-dorm-profile-2}
			}
		}
		\caption{Statistics of metaverse activities in a University network --  (a) spans of tracked flows toward primary and time-critical activity domains and (b) upstream packet rates of our ground-truth metaverse VR sessions; (c) upstream packet rates of VRChat primary and time-critical activity flows from dormitory residents during a long-weekend holiday (Saturday, Sunday, and Monday), (b) the volume consumed, and (e) time spent in the two available activity states (\ie HS and SUE).}
		\label{fig:field-6Sep}
	\end{center}
\end{figure*}

\subsubsection{Our ground-truth sessions}\label{sec:DeploymentGroundTruthSession}
Two of the authors who stay in the university dormitory (served by campus WiFi gateways) immersed themselves in the four studied metaverses for five consecutive days over a week. Each metaverse session was played once a day and lasted for over twenty minutes. In each session, the authors attended all user activity states available in the respective metaverse platform, encompassing events such as social meetups, multiplayer games, street talks and concerts. Three non-metaverse VR games were also played in the five days as comparisons.
After comparing our system logs, all our sessions in the four metaverses were successfully detected and classified with accurate temporal coverage, while other VR game sessions were not misidentified. The representative statistics of a two-hour period are shown in Fig.~\ref{fig:field-6Sep}, including the span of detected primary and time-critical activity flows in Fig.~\ref{fig:fieldFlow-6Sep} and their upstream packet rates in Fig.~\ref{fig:fieldPkt-6Sep}.

\subsubsection{Metaverse activity in a dormitory over a long-weekend holiday}\label{sec:DeploymentDormUsers}
During a long weekend from Saturday midnight to the next Tuesday, our real-time system detected ten metaverse sessions in the dormitory VLAN. 
They are all from one campus WiFi gateway serving a range of dormitory residents that heavily immersed in VRChat during the night (\ie from 12 AM to 4 PM) on Saturday and Sunday. In contrast, the metaverse sessions during the daytime of Monday were relatively short -- perhaps students had classes after the holiday on the following Tuesday. 

\begin{wraptable}{L}{7.5cm}
	\caption{User-perceived latency of each AS involved in VRChat by dormitory residents (shown by number of flows and their fractions).}
	\small
	\begin{tabular}{|l|l|l|l|l|}
		\hline
		\rowcolor[rgb]{ .906,  .902,  .902} 	\textbf{AS}       & \textbf{<10ms}   & \textbf{10 -- 20ms}  & \textbf{20 -- 50ms} & \textbf{>50ms}\\ \hline
		AS1 & 0 (0\%)&3724 (98.7\%) & 47 (1.3\%) & 0 (0\%) \\ \hline 
		AS2  & 1 (0.0\%) & 1819 (99.1\%) & 13 (0.7\%)& 1 (0.0\%) \\ \hline
		AS3  & 0 (0\%) & 0 (0\%) &0 (0\%) & 6 (100\%) \\ \hline
		AS4 & 0 (0\%) & 1535 (99.0\%)& 15 (1.0\%)& 0 (0\%) \\ \hline		
		AS5 & 0 (0\%) & 249 (98.8\%)& 3 (0.2\%)& 0 (0\%)\\ \hline
	\end{tabular}
	\label{tab:latencyDeployment}
\end{wraptable}

\textbf{Volume and time usage:} In Fig.~\ref{fig:fieldPkt-dorm}, we can visually see the upstream packet rates of the entire university network (``All''), primary flows (``VRC. Primary''), and time-critical activity flows (``VRC. Time-critical.''), respectively.
Fig.~\ref{fig:fieldPkt-dorm-profile-1} and Fig.~\ref{fig:fieldPkt-dorm-profile-2} show the volume consumed in Gigabyte and time spent in hour by the dorm metaverse sessions in the two available activity states of VRChat (\ie HS and SUE) over the long weekend. We can see that one-third user time were spent in the home space, which only consumed about one-sixth of the total upstream volume. This is aligned with our analysis that SUE is very demanding and require prioritized network resource provisioning by an operator for premium metaverse experience.

\textbf{Accessed ASes and latencies:} There are tens of ASes being accessed by the metaverse sessions detected in the wild. We now show five ASes serving primary domain content as representative examples. The per-flow latency from each AS is shown in Table~\ref{tab:latencyDeployment}. Each column represents a latency range. It is clear that most of the flows in AS1, AS2, AS4, and AS5 had good latencies (\ie smaller than 20ms), while a minor fraction of them were suffering (\ie 20 -- 50ms). Noted that, six flows toward AS3 that operates some primary domain services of VRChat were with bad latencies greater than 50ms, which possibly require routing/caching optimization by the respective operator.

While there is just a few metaverse session happens in our university campus network, our deployment demonstrates that \textit{MetaVRadar} provides operators visibility into metaverse sessions and their activity states in their networks. In addition, it helps them to understand the distribution of metaverse service domains and latencies seen by users as an important reference for experience-driven network optimizations in the near future.

\subsection{Limitations and Operational Considerations}\label{sec:discussionOnLimitation}
We acknowledge that our study focused on only four metaverse applications on the Oculus VR platform accessed via the Quest 2 headset. Variations may occur in emerging metaverse applications (\eg Microsoft's Mesh \cite{Mesh}) and on different VR platforms (\eg Apple's Vision Pro \cite{AppleVisionPro}). Additionally, our network measurements were conducted in our lab setup, which was served by an educational broadband network. It is important to note that network traffic characteristics may vary when using VR headsets connected to mobile wireless networks (\eg 4G and 5G).

When deploying \textit{MetaVRadar} in an operational ISP network to obtain customer insights on metaverse applications, there are three practical considerations.
Firstly, our proof-of-concept prototype covers four popular metaverse VR applications in our specific geographic region. However, for deployment in different contexts, the coverage will need to be reassessed and adjusted accordingly. To stay updated with the evolving market, it is a common industrial practice \cite{MLOps} for operators to train new models using our training pipeline\footnote{Source code for the training pipeline is available at \textit{https://github.com/RahulTripathi0401/MetaVRadar}.} for the inclusion of emerging metaverses.
Secondly, for the metaverse applications already included in the system, it is important to periodically re-train the detection signatures and user activity state classification models to accurately capture variations introduced by new releases/versions that may impact traffic characteristics.
Thirdly, we acknowledge the limitations in obtaining the training dataset for this study, such as manual data collection by authors playing metaverse VR applications and limited coverage of user activities. As a result, the trained detection signatures and classification models may suffer from the over-fitting problem. To address this issue in commercial deployments, operators of the customer insights platform can train the inference models using a dataset with larger volume and more comprehensive coverage of user activities compared to our research prototype.

\section{Related Work}\label{sec:relatedWork}
Many prior works analyze network traffic characteristics of specific networked applications for detection and performance optimization, such as streaming video \cite{HHGharakheiliTNSM2019}, 360-degree video \cite{TXuCoNEXT2019}, online conferencing \cite{KMacMillanICM2021,HChangIMC2021}, WiFi voice calls \cite{SCMadanapalliLCN2019}, TLS usage of IoT devices \cite{MTalhaParachaIMC2021}, and online games \cite{AWongRASSE2021, SCMadanapalliPAM2022}. The work in \cite{SCMadanapalliPAM2022} focuses on detecting gameplay on PC platforms by analyzing packet byte signatures in gaming UDP flows, which serves as an inspiration for our method of detecting time-critical activity domain flows. Those works provide insights into the network behaviours of their studied applications, helping operators better understand their networks' utilization and spot potential operational issues. 
 
In the context of metaverse VR applications, apart from many studies on its enabling technologies \cite{HNingArXiv2021,QYangArXiv2022,LLeeArXiv2021,HDuanMM2021} such as AR/VR \cite{OAbariNSDI2017,JBrendanTVCG2020,ZLaiTMC2020,JLiISMAR2021,LLiuMobiSys2018,JMengASPLOS2020,SShiMobiSys2019}, motion sensing \cite{KAhujaCHI2022}, and platform governance \cite{LBlackwellCSCW2019,KBoosMobiSys2016,FGuoHCI2022}, their architectural problems and scalability issues to be solved by metaverse developers (\eg bandwidth consumptions with increasing number of users, CPU/GPU utilization of terminal device, and rendering cost of user avatars) have been studied in \cite{MetaverseQoE,cheng2022are}. The authors of \cite{cheng2022are} have also pointed out that one critical scalability issue that could be solved by changing the operational architecture of metaverses is that the consumptions of user bandwidth and CPU/GPU resources currently depend on the number of active users, thus, can introduce high overhead on the VR headset -- one potential solution is remote rendering, \ie rendering application graphics on the cloud server instead of on user VR devices.

In this paper, we focus on analyzing the network anatomy (\eg utilization of service domains, flows, and packets) of metaverse VR applications for effective detection and state classification of user activity to help network operator get ``metaverse ready'', instead of performance and scalability issues to be addressed by metaverse developers as studied in prior works \cite{MetaverseQoE,cheng2022are}. Insights from this paper serve as preliminary knowledge to the future works that prepare network readiness for emerging metaverse VR applications.

\section{Conclusion}\label{sec:conclusion}
Our work has provided a detailed anatomy of metaverse VR network traffic, related it to user activity, and associated it with user experience. We first categorized user activity into six distinct states, and identified unique traffic characteristics associated with these states. We then designed and implemented \textit{MetaVRadar}, a three-stage method to detect metaverse activity and classify user state by extracting attributes on flow behavior and volumetric patterns via real-time network traffic analysis. Evaluation of our prototype in the lab shows promising results across four popular metaverse applications on the Oculus VR platform, and evaluation in the field showed that user-generated content plays a dominant role in user experience. Our study is a first step towards helping telecommunications network operators make their infrastructure ``metaverse-ready''.

\section*{Acknowledgement}
We thank our shepherd Niklas Carlsson and the anonymous reviewers for their insightful feedback. This work is supported by research funding from Canopus Networks.

\bibliographystyle{ACM-Reference-Format}
\bibliography{Reference}

\received{February 2023}
\received[revised]{October 2023}
\received[accepted]{October 2023}

\appendix

\pagebreak

\section{Detection signatures of time-critical activity domain flows}\label{sec:timecriticalDomainByteSignature}
Example sequences of packet payload size for time-critical activity domain flows are given in Table~\ref{tab:timecriticalDomainByteSignature}. Each per-flow signature is listed as an individual array enclosed by square bracket after its respective service UDP port number. For example, the first item in Table~\ref{tab:timecriticalDomainByteSignature} for Multiverse is ``\textbf{5055}: [56,86,32,143]'', indicating that a UDP flow from a tracked Multiverse user to the service port 5055 with its first four packets sized 56, 86, 32, and 143 will be detected as a time-critical activity domain flow.

\begin{table*}[h!]
	\centering
	\caption{UDP ports and upstream packet payload size sequences of flows toward time-critical activity domains.}
	\small
	\begin{tabular}{|l|c|}
		\hline
		\rowcolor[rgb]{ .906,  .902,  .902} 	\textbf{Metaverse}     & \textbf{Example payload size sequence signatures towards each port} \\ \hline
		Multiverse   &  \textbf{5055}: [56,86,32,143], \textbf{5056}: [56,56,85,159], \textbf{5058}: [56,86,159,174]\\ \hline
		VRChat     & \textbf{5055}: [60,60,34,69], \textbf{5056}: [60,89,163,36,1200], \textbf{5058}: [60,89,163,68]   \\ \hline
		Rec Room      & \textbf{5055}: [13,13,13,13,13], \textbf{5056}: [56,32,65,65], \textbf{5058}: [56,32,65,32] \\ \hline
		AltSpaceVR   & \textbf{5055}: [56,85,32,143,32,1196], \textbf{5056}: [56,85,163,44,1196]\\ \hline
	\end{tabular}
	\label{tab:timecriticalDomainByteSignature}
\end{table*}

\section{Training and Evaluating Classifiers for User Activity States}\label{sec:AppendixTrainingDetails}
This section presents important technical details of the training and evaluation of classifiers for user activity states, which are not covered in the main body of this paper. We discuss the full list of volumetric attributes per metaverse session (\S\ref{sec:AppendixAttributeList}), training process (\S\ref{sec:AppendixTrainingProcess}), performance of baseline stateless classifiers (\S\ref{sec:AppendixClassification}), and evaluation of our classification method using open set (\S\ref{sec:AppendixOpenSet}).

\subsection{Full List of Classification Attributes}\label{sec:AppendixAttributeList}
We obtained forty volumetric attributes to describe network traffic characteristics of a metaverse user session using the methods defined in Fig.~\ref{fig:TrafficAttribute}. Those attributes are computed per monitoring interval for both metaverse domain types (\ie primary and time-critical activity) including flow-level statistics (\ie per flow volume, packet count, and average packet size) represented by median and standard deviation (\ie $\sigma$) and host-level statistics including number of flows, volume, and packet count. 
A full list of attributes is provided in Table~\ref{tab:fullAttributeList} with their labels, full names, and short descriptions.

\begin{table}[h!]
	\centering
	\caption{Full List of volumetric attributes for metaverse user activity classification as defined in Fig.~\ref{fig:TrafficAttribute}.}
	\label{tab:fullAttributeList}
	\small
	\begin{tabular}{|c|l|l|}
		\hline
		\rowcolor[rgb]{ .906,  .902,  .902} 	\textbf{Label} & \multicolumn{1}{c|}{\textbf{Attribute name}}                 & \multicolumn{1}{c|}{\textbf{Description (per monitoring interval)}}                                         \\ \hline
		\textbf{A1}        & tcp\_prim\_mdn\_vol                              & The median of volume per primary TCP flow                              \\ \hline
		\textbf{A2}        & tcp\_prim\_mdn\_pkt\_ct                          & The median of packet count per primary TCP flow                               \\ \hline
		\textbf{A3}        & tcp\_prim\_mdn\_pkt\_sz                          & The median of average packet size per primary TCP flow                         \\ \hline
		\textbf{A4}        & tcp\_prim\_std\_vol                              & The $\sigma$ of volume per primary TCP flow                      \\ \hline
		\textbf{A5}        & tcp\_prim\_std\_pkt\_ct                          & The $\sigma$ of packet count per primary TCP flow                   \\ \hline
		\textbf{A6}        & tcp\_prim\_std\_pkt\_sz                          & The $\sigma$ of average packet size per primary TCP flow            \\ \hline
		\textbf{A7}        & tcp\_actv\_mdn\_vol             & The median of packet volume per time-critical actv. TCP flow               \\ \hline
		\textbf{A8}        & tcp\_actv\_mdn\_pkt\_ct         & The median of packet count per time-critical actv. TCP flow                 \\ \hline
		\textbf{A9}        & tcp\_actv\_mdn\_pkt\_sz         & The median of average packet size per time critical-actv. TCP flow            \\ \hline
		\textbf{A10}       & tcp\_actv\_std\_vol             & The $\sigma$ of volume per time-critical actv. TCP flow          \\ \hline
		\textbf{A11}       & tcp\_actv\_std\_pkt\_ct         & The $\sigma$ of packet count per time-critical actv. TCP flow          \\ \hline
		\textbf{A12}       & tcp\_actv\_std\_pkt\_sz         & The $\sigma$ of average packet size per time-critical actv. TCP flow \\ \hline
		\textbf{A13}       & udp\_prim\_mdn\_vol                              & The median of volume per primary UDP flow                               \\ \hline
		\textbf{A14}       & udp\_prim\_mdn\_pkt\_ct                          & The median of packet count per primary UDP flow                             \\ \hline
		\textbf{A15}       & udp\_prim\_mdn\_pkt\_sz                          & The median of of average packet size per primary UDP flow                         \\ \hline
		\textbf{A16}       & udp\_prim\_std\_vol                              & The $\sigma$ of volume per primary UDP flow                  \\ \hline
		\textbf{A17}       & udp\_prim\_std\_pkt\_ct                          & The $\sigma$ of packet count per primary UDP flow                     \\ \hline
		\textbf{A18}       & udp\_prim\_std\_pkt\_sz                          & The $\sigma$ of average packet size per primary UDP flow           \\ \hline
		\textbf{A19}       & udp\_actv\_mdn\_vol             & The median of volume per time-critical actv. UDP flow               \\ \hline
		\textbf{A20}       & udp\_actv\_mdn\_pkt\_ct         & The median of packet count per time-critical actv. UDP flow                \\ \hline
		\textbf{A21}       & udp\_actv\_mdn\_pkt\_sz         & The median of average packet size per time-critical actv. UDP flow           \\ \hline
		\textbf{A22}       & udp\_actv\_std\_vol             & The $\sigma$ of volume per time-critical actv. UDP flow         \\ \hline
		\textbf{A23}       & udp\_actv\_std\_pkt\_ct         & The $\sigma$ of packet count per time-critical actv. UDP flow         \\ \hline
		\textbf{A24}       & udp\_actv\_std\_pkt\_sz         & The $\sigma$ of average packet size per time-critical actv. UDP flow \\ \hline
		\textbf{A25}       & tcp\_prim\_\#\_cncr\_flow                  & The number of concurrent TCP primary flows                       \\ \hline
		\textbf{A26}       & tcp\_prim\_\#\_new\_flow                         & The number of new TCP primary flows                 \\ \hline
		\textbf{A27}       & tcp\_prim\_vol                                   & The TCP volume to primary domain                                         \\ \hline
		\textbf{A28}       & tcp\_prim\_pkt\_ct                               & The TCP packet count to primary domain                                   \\ \hline
		\textbf{A29}       & tcp\_actv\_\#\_cncr\_flow & The number of concurrent TCP time-critical actv. flows        \\ \hline
		\textbf{A30}       & tcp\_actv\_\#\_new\_flow        & The number of new TCP time-critical actv. flows   \\ \hline
		\textbf{A31}       & tcp\_actv\_vol                  & The TCP volume to time-critical actv. domain                        \\ \hline
		\textbf{A32}       & tcp\_actv\_pkt\_ct              & The TCP packet count to time-critical actv. domain                    \\ \hline
		\textbf{A33}       & udp\_prim\_\#\_cncr\_flow                  & The number of concurrent UDP primary flows                       \\ \hline
		\textbf{A34}       & udp\_prim\_\#\_new\_flow                         & The number of new UDP primary flows                   \\ \hline
		\textbf{A35}       & udp\_prim\_vol                                   & The UDP volume to primary domain                                         \\ \hline
		\textbf{A36}       & udp\_prim\_pkt\_ct                               & The UDP packet count to primary domain                                   \\ \hline
		\textbf{A37}       & udp\_actv\_\#\_cncr\_flow & The number of concurrent UDP time-critical actv. flows         \\ \hline
		\textbf{A38}       & udp\_actv\_\#\_new\_flow        & The number of new UDP time-critical actv. flows     \\ \hline
		\textbf{A39}       & udp\_actv\_vol                  & The UDP volume to time-critical actv. flows                          \\ \hline
		\textbf{A40}       & udp\_actv\_pkt\_ct              & The UDP packet count to time-critical actv. flows                    \\ \hline
	\end{tabular}
\end{table}

\subsection{Training Process}\label{sec:AppendixTrainingProcess}
In our design, each metaverse application has its own user activity state classifiers to effectively capture variations in volumetric profiles across different applications.

Initially, we trained stateless machine-learning (ML) classifiers (represented by the blue block in Fig.\ref{fig:classification}) that utilize volumetric attributes of a metaverse session per inference interval. Subsequently, we employed a stateful classification approach (represented by the green, yellow, and grey blocks in Fig.\ref{fig:classification}) which incorporates volumetric attributes, past user activity states, and classification confidence.
To achieve optimal classification performance, we tuned available parameters in our training processes, including the underlying algorithms of machine learning models, model parameters, monitoring intervals, attribute types, the number of past user activity states, and thresholds on classification confidence.

As the evaluation metric, we use accuracy in the format of ``true positive (TP) | false positive (FP)'' generated by ten-fold validation method using our lab training dataset. 
After each training stage, we selected the classifiers that achieved the highest accuracies for each metaverse application, which were then subjected to further evaluations using an open set collected within our lab network and ground-truth sessions in a different network environment, specifically the campus dormitory.

Now we provide training details of both stateless and stateful classification.

\subsubsection{Stateless ML Classifiers}\label{sec:AppendixTrainingStateess}
To build the stateless ML classifiers (\ie the blue block in Fig.~\ref{fig:classification}), we compared two popular machine-learning algorithms available in the \textit{scikit-learn} library, \ie neural networks and random forest. The classifiers are iteratively trained with their respective tunable parameters to achieve the best classification accuracies. 

For random forest classifiers, the number of trees, tree depth, and the maximum number of attributes per tree are tuned for the ranges of [5, 300], [1, 16], and [2, 40], respectively. 
Classifiers built on neural networks were trained using different solver functions (either limited-memory BFGS, stochastic gradient descent, or Adam function), the number of nodes per layer (from 5 to 500), and the number of hidden layers (from 5 to 500). 
We note that, for our dataset, the best accuracies achieved by neural networks for all metaverse applications are noticeably lower than (\ie more than 5\%) those by random forest models, thus, are not further considered in our implementation.

\subsubsection{Stateful Classification Parameters}
As discussed in \S\ref{sec:thirdStage}, the stateful classification approach depicted in Fig.~\ref{fig:classification} incorporates a stateful classifier that uses volumetric attributes and past \textit{N} states to infer the current user activity state. If the confidence level of an inference result from the stateful classifier falls below a threshold \textit{T}, the current result will rely on the stateless classifier.

The training process for stateful classifiers is similar to that for stateless classifiers, as discussed in \S\ref{sec:AppendixTrainingStateess}. However, there is an additional parameter to consider, \ie the number of past states \textit{N}. It is tuned from 1 to 10. The best accuracies for all metaverse applications are achieved with \textit{N} set to 5. 
Regarding the confidence threshold \textit{T}, it is tuned from 30\% to 95\%, and the best accuracies are attained at 80\%.

\subsection{Baseline Evaluation with Stateless Classifiers}\label{sec:AppendixClassification}
As a baseline evaluation of our stateful classification approach discussed in \S\ref{sec:thirdStage}, we now report the accuracy of the classifications that are only with the stateless ML classifiers (\ie the blue block in Fig.~\ref{fig:classification}) applied to volumetric attributes per metaverse session.

As shown in Fig.~\ref{fig:TrafficAttribute}, there are two types of volumetric attributes of each metaverse session, \ie host-level and flow-level statistics.
Host-level attributes are relatively lightweight for real-time computation in terms of CPU time and RAM usage, whereas flow-level attributes require more computational resource for a finer-grained visibility. 
In the following evaluation, we assess the baseline accuracy of stateless classifications using only host-level statistics and both host- and flow-level attributes. It is important to note that each metaverse may not support certain user activity states, which are indicated as `` -- '' in the respective tables.

\subsubsection{Host-level Attributes}
We now discuss the accuracy of stateless classifiers using only host-level attributes, which were trained for each metaverse application studied in this paper. The classification accuracy, presented in Table~\ref{tab:classificationAccuracyStatelessOnlyHostLevel}, is provided in the format of ``TP|FP'', consistent with the format used for our stateful classification approach shown in Table~\ref{tab:classificationAccuracyStateful}.

From the results, it is evident that VRChat demonstrates decent accuracies, with true positive (TP) rates exceeding 90\%. This may be attributed to the presence of only two available activity types in VRChat. However, both Multiverse and Rec Room exhibit lower TP rates below 50\% and higher false positive (FP) rates exceeding 30\% for certain state types such as HS and CC. 
As for the results using Neural Networks models, the best TP|FP for VRChat, Multiverse and Rec Room are 90\%|9.3\%, 51\%|11\% and 24\%|34\%, respectively, which are much lower than the best results using Random Forest models. Therefore, Neural Networks models are not further considered in our improved classifications using flow-level attributes and stateful approach.
It is clear that when considering only host-level volumetric attributes, accurately classifying user activity states in most metaverse applications is challenging.

\begin{table}[!h]
	\caption{Classification accuracy (\ie TP|FP) of user activity states using our \textbf{stateless machine-learning classifiers} with \textbf{only host-level statistics} on the four metaverse VR applications.}
	\label{tab:classificationAccuracyStatelessOnlyHostLevel}
	\begin{tabular}{|l|l|l|l|l|l|l|}
		\hline
		\rowcolor[rgb]{ .906,  .902,  .902} \textbf{Metaverse} & \textbf{\color{red}HS}        & \textbf{\color{blue}MH} & \textbf{\color{purple}SUE} & \textbf{\color{olive}SPE} & \textbf{\color{orange}CC} & \textbf{\color{brown}AT} \\ \hline
		Multiverse & 20\%|0.7\% & 83\%|9.5\%   &   81\%|9.0\%  &  62\%|3.3\%   & -- & 79\%|21\%   \\ \hline
		VRChat       &     92\%|1.1\%        & --   &  97\%|1.6\%   & --  & -- & -- \\ \hline
		Rec Room     &        36\%|0.9\%    & 84\%|1.7\%   &  88\%|4.0\%   & --    &  95\%|35\%  & 25\%|1.2\%		\\ \hline
		AltSpaceVR   &  96\%|1.0\%  & 76\%|5.0\%   &   86\%|9.9\%  & --  & -- & -- \\ \hline

	\end{tabular}
\end{table}

\subsubsection{Host- and Flow-level Attributes}
We have also provided the accuracy of stateless classifiers that incorporate both host- and flow-level attributes in Table~\ref{tab:classificationAccuracyStateless}. These results demonstrate significant improvements for several states that could not be accurately classified using only host-level attributes.
For instance, in Multiverse, the true positive (TP) rate for the HS state has improved from 20\% to 100\%, and in Rec Room, the false positive (FP) rate for the CC state has decreased from 35\% to 6.5\%.

However, there are still certain states (SPE, AT, and MH) that are not being accurately classified, with TP rates ranging from 66\% to 82\%, which also results in high FP rates for other states.
As discussed in \S\ref{sec:Development}, these three states do not consistently exhibit distinct traffic patterns, making them better captured by our stateful classification approach (discussed in \S\ref{sec:thirdStage}). Our stateful approach achieves TP rates of over 90\% and FP rates below 2\% for all states.

\begin{table}[!h]
	\caption{Classification accuracy (\ie TP|FP) of user activity states using our \textbf{stateless machine-learning classifiers} with \textbf{both host- and flow-level statistics} on the four metaverse VR applications.}
	\label{tab:classificationAccuracyStateless}
	\begin{tabular}{|l|l|l|l|l|l|l|}
		\hline
		\rowcolor[rgb]{ .906,  .902,  .902} \textbf{Metaverse} & \textbf{\color{red}HS}        & \textbf{\color{blue}MH} & \textbf{\color{purple}SUE} & \textbf{\color{olive}SPE} & \textbf{\color{orange}CC} & \textbf{\color{brown}AT} \\ \hline
		Multiverse & 100\%|0.7\% & 82\%|5.8\%   &   93\%|5.0\%  &  78\%|4.1\%   & -- & 85\%|0\%   \\ \hline
		VRChat       &     95\%|0\%        & --   &  100\%|5.4\%   & --  & -- & -- \\ \hline
		Rec Room     &        92\%| 5.1\%    & 90\%|1.2\%   &  92\%|3.8\%   & --    &  91\%|6.5\%  & 66\%|0.6\%	\\ \hline
		AltSpaceVR   &  100\%|0\%  & 81\%|4.0\%   &   88\%|6.4\%  & --  & -- & -- \\ \hline
	\end{tabular}
\end{table}

\subsection{Open Set Evaluation}\label{sec:AppendixOpenSet}
We further evaluated the detection and classification accuracy of our \textit{MetaVRadar} prototype by assessing its performance using an open set of data. The open set was not exposed during the training process and was collected six months after the training dataset was gathered. This ensures that the open set contained a previously unseen distribution of samples, allowing us to evaluate the robustness of our approach.
Unfortunately, Microsoft shut down AltspaceVR to prioritize the development of its coming metaverse platform, Mesh \cite{Mesh}, before we collected the open set. Therefore, AltspaceVR is not included in this evaluation.
For each of the three considered metaverse VR applications, we collected eight sessions, each encompassing all available user activity states and lasting for at least 30 minutes.

\begin{table}[!h]
	\caption{Classification accuracy obtained from the \textbf{open set} of three metaverse VR applications (excluding AltspaceVR) collected after six months of training process.}
	\label{tab:classificationAccuracyOpenSet}
	\begin{tabular}{|l|l|l|l|l|l|l|}
		\hline
		\rowcolor[rgb]{ .906,  .902,  .902} \textbf{Metaverse} & \textbf{\color{red}HS}        & \textbf{\color{blue}MH} & \textbf{\color{purple}SUE} & \textbf{\color{olive}SPE} & \textbf{\color{orange}CC} & \textbf{\color{brown}AT} \\ \hline
		Multiverse & 98\%|1.3\% & 90\%|4.5\%   &   93\%|3.3\%  &  84\%|2.5\%   & -- & 84\%|1.3\%   \\ \hline
		VRChat       &     83\%|2.7\%        & --   &  97\%|16.4\%   & --  & -- & -- \\ \hline
		Rec Room     &        94\%|6.3\%    & 87\%|2.8\%   &  83\%|0.8\%   & -- &  80\%|2.8\%  & 94\%|3.0\%
		\\ \hline
	\end{tabular}
\end{table}

In the open set evaluation, we achieved a 100\% detection accuracy, successfully identifying all metaverse user sessions. However, as shown in Table~\ref{tab:classificationAccuracyOpenSet}, we observed a slight decrease in the accuracy of activity state classification. Specifically, the accuracy dropped by approximately 5\% to 10\% for all three metaverses. This decrease can be attributed to the training dataset that becomes slightly outdated as a result of several application updates occurred over the six-month period.
We note that after re-training models with newly collected datasets (\ie training on four datasets and testing on the other four datasets), the overall accuracy became high again, exceeding 95\% true positive and below 5\% false positive rates for all metaverse applications. This observation emphasizes the need for periodic model re-training in commercial deployment as discussed in \S\ref{sec:discussionOnLimitation}.

\end{document}